\pdfoutput=1
\providecommand{\BuildMode}{combined}
\documentclass[letterpaper,twocolumn,10pt]{article}
\usepackage{etoolbox}
\usepackage{xr-hyper}
\usepackage{placeins}
\usepackage{usenix2019_v3}

\usepackage{graphicx}
\usepackage{booktabs}
\usepackage{array}
\usepackage{amsmath,amssymb,amsthm}
\usepackage{xspace}
\usepackage{caption}
\usepackage{subcaption}
\usepackage{tikz}
\usepackage{pgfplots}
\pgfplotsset{compat=1.18}
\definecolor{fshit}{HTML}{D55E00}
\definecolor{fsmiss}{HTML}{0072B2}
\definecolor{fsexec}{HTML}{009E73}
\definecolor{fspred}{HTML}{882255}
\definecolor{fsshared}{HTML}{000000}
\definecolor{fsiso}{HTML}{6B6B6B}
\definecolor{fsink}{HTML}{000000}
\definecolor{fsref}{HTML}{999999}
\definecolor{fsother}{HTML}{969696}
\definecolor{fspale}{HTML}{F2F2F2}
\definecolor{fshatch}{HTML}{6B6B6B}
\definecolor{fshittint}{HTML}{F4D7BF}
\definecolor{fsmisstint}{HTML}{BFDCEC}
\definecolor{fsexectint}{HTML}{BFE7DC}
\definecolor{fspredtint}{HTML}{E1C8D4}
\definecolor{fssharedtint}{HTML}{D9D9D9}
\definecolor{fsisotint}{HTML}{D9D9D9}
\definecolor{fswhite}{HTML}{FFFFFF}
\definecolor{fsleak1}{HTML}{F7DFCC}
\definecolor{fsleak2}{HTML}{ECB78C}
\definecolor{fsleak3}{HTML}{E18B47}
\definecolor{fsleak4}{HTML}{D55E00}
\definecolor{fsleak5}{HTML}{8A3D00}

\usetikzlibrary{arrows.meta,calc,positioning,patterns,patterns.meta,decorations.pathreplacing,fit}
\tikzset{
  fs/fig/.style={font=\footnotesize, line width=0.6pt, line cap=butt, line join=miter,
    inner sep=1.5pt, outer sep=0pt, >={Stealth[length=3pt,width=2.4pt]}},
  fs/thin/.style={line width=0.4pt},
  fs/thick/.style={line width=1pt},
  fs/label/.style={inner sep=1pt, text=fsink},
  fs/panel/.style={fs/label, anchor=north west, font=\footnotesize\bfseries},
  fs/world/.style={fs/label, anchor=east},
  fs/box/.style={draw=fsink, fs/thin, fill=fswhite, rectangle, align=center, minimum height=5mm},
  fs/cell/.style={draw=fsink, fs/thin, fill=fswhite, rectangle, minimum width=2.5mm, minimum height=3mm},
  fs/step/.style={draw=fsink, fs/thin, fill=fswhite, rectangle},
  fs/req/.style={draw=fsink, fs/thin, fill=fswhite, rectangle},
  fs/other/.style={draw=fsink, fs/thin, fill=fsother, rectangle},
  fs/hit/.style={draw=fshit, fs/thin, fill=fshittint, rectangle},
  fs/miss/.style={draw=fsmiss, fs/thin, fill=fsmisstint, rectangle},
  fs/saved/.style={draw=fshit, fs/thin, dash pattern=on 1.5pt off 1pt, rectangle,
    pattern={Lines[angle=45, distance=2pt, line width=0.5pt]}, pattern color=fshit},
  fs/slot/.style={draw=fsink, fs/thin, dash pattern=on 1.5pt off 1pt, fill=fswhite, rectangle},
  fs/fab/.style={draw=fsink, fs/thin, rectangle,
    pattern={Dots[distance=1.6pt, radius=0.4pt]}, pattern color=fshatch},
  fs/cached/.style={draw=fsink, fs/thin, fill=fspale, rectangle},
  fs/recomp/.style={draw=fsink, fs/thin, rectangle,
    pattern={Lines[angle=-45, distance=2pt, line width=0.5pt]}, pattern color=fshatch},
  fs/later/.style={draw=fsink, fs/thin, dash pattern=on 0.6pt off 1pt, rectangle},
  fs/arrow/.style={draw=fsink, fs/thin, ->},
  fs/flow/.style={draw=fsink, line width=0.6pt, ->},
  fs/time/.style={draw=fsink, fs/thin, ->},
  fs/shift/.style={draw=fshit, fs/thick, {Bar[width=4pt]}-{Stealth[length=3pt,width=2.4pt]}},
  fs/brace/.style={draw=fsink, fs/thin, decorate, decoration={brace, amplitude=2pt, raise=1pt}},
  fs/ref/.style={draw=fsref, fs/thin, dash pattern=on 2pt off 1.5pt},
  fs/release/.style={draw=fsink, fs/thick},
  fs/dist hit/.style={draw=fshit, line width=1pt},
  fs/dist miss/.style={draw=fsmiss, line width=1pt, dash pattern=on 3pt off 1.5pt},
  fs/dist target/.style={draw=fsink, line width=1pt},
  fs/dist unprotected/.style={draw=fshit, line width=0.8pt, dash pattern=on 0.6pt off 1.2pt},
  fs/dist fill/.style={fill=fspale, draw=none},
}

\graphicspath{{figures/}}

\newcommand{\SysNamePlain}{Isonomy}
\newcommand{\PaperSubtitle}{Isolating Every Request from Timing Side Channels at Nearly Full-Sharing Throughput in Multi-Instance LLM Prefill}

\hypersetup{pdfauthor={Bowen Sun}, pdftitle={\SysNamePlain: \PaperSubtitle}}

\newcommand{\sys}{\SysNamePlain\xspace}
\newcommand{\PaperTitle}{\SysNamePlain: \PaperSubtitle}

\newcommand{\eps}{\varepsilon}

\theoremstyle{plain}
\newtheorem{theorem}{Theorem}
\newtheorem{proposition}{Proposition}
\newtheorem{lemma}{Lemma}
\theoremstyle{definition}
\newtheorem{definition}{Definition}

\newif\ifIncludeMain
\newif\ifIncludeSupplement
\ifdefstring{\BuildMode}{main}{\IncludeMaintrue}{%
  \ifdefstring{\BuildMode}{supplement}{\IncludeSupplementtrue}{%
    \ifdefstring{\BuildMode}{combined}{\IncludeMaintrue\IncludeSupplementtrue}{%
      \PackageError{manuscript}{Unknown build mode '\BuildMode'}%
        {Set \string\BuildMode\space to main, supplement, or combined.}}}}

\ifIncludeMain
  \title{\PaperTitle}
  \ifIncludeSupplement\else
  \fi
\else
  \title{Supplementary Material for \\ \PaperTitle}
  \hypersetup{pdftitle={Supplementary Material for \PaperTitle}}
\fi
\author{
{\rm Bowen Sun}\\
Johns Hopkins University
\and
{\rm Chaowei Xiao}\\
Johns Hopkins University
}
\date{}

\begin{document}
\maketitle

\ifIncludeMain
  \begin{abstract}
  Requests to an LLM service often repeat a long prefix, such as an agent's instructions or a conversation
history, and reusing its cached keys and values saves prefill computation.
A cache hit, however, makes the response arrive earlier and can reveal that another request used the
same prefix, so providers share the cache only within an account.
Under least privilege, even permission to send and pay for requests should not expose other requests'
prompts in the same account.
This matters because agents send unreviewed requests carrying private data, resellers pool users behind
one account, and credentials can be stolen.
\sys therefore isolates each request by default and lets requests opt into explicit sharing groups.
Rather than give up reuse, it keeps one cache shared across accounts and hides only hits on prefixes
cached outside a request's group.
To hide such hits, prior work suggests that the savings from cache reuse must be hidden from every
request, not only from the one that hits.
However, our measurements on commercial APIs found no sign that a request can observe the work done
for others, so there it may suffice to delay only the request that hits.
Built on vLLM, \sys releases a response with such a hit only when it would have been ready as a miss, on
the instance that routing would then have chosen.
It either estimates that time and randomizes the release, or lets other requests fill the work that the
hit skips.
On eight instances, timing probes detect hits with an estimated advantage over guessing of at most
0.07, against up to 0.98 without protection.
\sys also retains up to 0.98 of unprotected sharing's throughput, against 0.95 for sharing within an
account, so stricter isolation need not cost performance.

  \end{abstract}

  \providecommand{\pend}{\textcolor{red}{\fbox{\phantom{0.00}}}}
\section{Introduction}\label{sec:intro}

Many prompts sent to an LLM service share a prefix, such as an agent's instructions and tool definitions, an
application's system prompt, a retrieved document, or a conversation history. Prefix caching reuses the
keys and values (KV) computed for such a prefix instead of computing them again~\cite{kwon2023efficient,zheng2024sglang}.
We study the prefill instances in a cluster, where this reuse is especially valuable: prefill-decode disaggregation gives
prefill its own pool of instances~\cite{zhong2024distserve,patel2024splitwise,qin2025mooncake}, and
cache-aware routing prefers an instance that already caches a request's prefix~\cite{srivatsa2025preble,llmd2026llmd,nvidia2026dynamo}.
A cache hit, however, also makes the response arrive earlier, so a \emph{probe}, a request with a guessed prefix, can
reveal through its timing that another request used the prefix, or even expose its
prompt~\cite{song2025early,zheng2026inputsnatch,wu2025know,wang2026optileak}.
After an audit found cross-user sharing on 7 of 17 commercial APIs~\cite{gu2025auditing}, providers
restricted sharing to an account, such as an organization, workspace, or project~\cite{openai2026prompt,anthropic2026prompt,google2026geminizdr}.
This boundary assumes that all requests in an account trust one another.

That trust is more than an API key should confer. A key authorizes its holder to send requests as a user
and charge them to an account, and under least privilege~\cite{saltzer1975protection} it should grant no
insight into other requests made with the same key. When a person writes every request and is the only one to read the responses, this distinction
matters little. An agent breaks both conditions: it sends requests without review, and it can disclose
what it learns through the network or the file system. Instructions injected through untrusted content
can thus make it probe other requests while it handles private data~\cite{debenedetti2025defeating,beurerkellner2025design}.
Account isolation also leaves the users of relays and resellers sharing one cache~\cite{fahey2026cacheprobe,ji2026uncovering},
and a stolen credential can expose its owner's prompts as well as incur charges~\cite{meli2019bad}.
We therefore isolate every request by default and let requests opt into sharing by naming a common group
within their account.

The simplest way to enforce this policy is to prevent a request from reusing anything cached outside its
sharing group. vLLM, SGLang, and Fireworks already let a client restrict reuse in this way, down to a single
request~\cite{vllm2026v026source,sglang2026radixcache,fireworks2026prompt}.
With every request isolated, however, this forfeits all reuse and much of the throughput it provides.
We instead let all requests share one cache, across groups and accounts. A request may then hit a prefix
cached outside its group, and we require only that its timing not reveal that hit.

\begin{figure*}[t]
\centering
\begin{tikzpicture}[fs/fig, x=1mm, y=1mm]
  \draw[fs/time] (13,0) -- (53,0);
  \node[fs/label, anchor=north east] at (53,-0.3) {time};
  \node[fs/label, anchor=north] at (16,-0.3) {arrival};
  \node[fs/world] at (12.3,4.75) {Miss twin};
  \draw[fs/miss] (13.5,2.5) rectangle (44,7);
  \draw[draw=fsmiss, fs/thin] (33,2.5) -- (33,7);
  \node[fs/label] at (23.25,4.75) {cached prefix};
  \node[fs/label] at (38.5,4.75) {rest};
  \node[fs/world] at (12.3,14.25) {Hit case};
  \draw[fs/hit] (13.5,12) rectangle (24.5,16.5);
  \node[fs/label] at (19,14.25) {rest};
  \draw[fs/ref] (24.5,14.25) -- (43.5,14.25);
  \node[fs/label, anchor=south] at (34,14.5) {held};
  \draw[fs/release] (44,11) -- (44,17.5);
  \node[fs/label, anchor=west] at (44.6,14.25) {release};
  \draw[fs/ref] (44,7) -- (44,11);
  \node[fs/box, anchor=south west, minimum width=39.5mm, inner sep=1.2pt] (rc) at (13.5,21) {release controller};
  \draw[fs/arrow] (16.5,16.5) -- (16.5,21);
  \draw[fs/arrow] (44,21) -- (44,17.8);
  \draw[fs/ref] (57.5,-6) rectangle (176.8,27.5);
  \draw[fs/ref] (rc.north east) -- (57.5,27.5);
  \draw[fs/ref] (rc.south east) -- (57.5,-6);
  \node[fs/label, anchor=north west] at (59,26.6) {Release controller};
  \node[fs/miss, minimum width=14mm, minimum height=4.5mm, inner sep=0pt] (tw) at (68,18.6) {miss twin};
  \foreach \x in {61, 76} {
    \draw[fs/thin, draw=fsink, fill=fswhite] (\x,-1) rectangle (\x+12,11);
  }
  \draw[fs/thin, draw=fsmiss, fill=fswhite] (91,-1) rectangle (103,11);
  \foreach \x/\k in {61/0, 61/1, 76/0, 76/1, 76/2, 91/0} {
    \draw[fs/other] (\x+1.5,-0.3+2.6*\k) rectangle (\x+10.5,1.8+2.6*\k);
  }
  \draw[fs/miss] (92.5,2.3) rectangle (101.5,4.4);
  \draw[draw=fshit, fill=fshittint, fs/thin, dash pattern=on 1.5pt off 1pt] (62.5,7.7) rectangle (71.5,10.3);
  \node[fs/label, anchor=south west] at (60.6,11.4) {hidden prefix};
  \node[fs/label, anchor=north] at (82,-1.8) {instances};
  \draw[fs/arrow] (tw.east) -| (97,11);
  \node[fs/label, anchor=south west] at (76.5,19) {routing rule (\S\ref{sec:routing})};
  \draw[fs/ref] (103,11) -- (115.7,26.5);
  \draw[fs/ref] (103,-1) -- (115.7,-5);
  \node[fs/label] at (111,10.75) {either};
  \begin{scope}[xshift=-1.6mm]
  \draw[fs/thin, draw=fsink, fill=fswhite] (117.3,11.3) rectangle (177.3,26.5);
  \draw[fs/thin, draw=fsink, fill=fswhite] (117.3,-5) rectangle (177.3,10.2);
  \node[fs/label, anchor=north west] at (118.2,25.9) {Prediction mode (\S\ref{sec:pred})};
  \node[fs/label, anchor=east] at (125.8,17.15) {step};
  \draw[fs/step] (126.5,12.4) rectangle (152,21.9);
  \draw[fs/other] (127.3,17.6) rectangle (135.3,21.1);
  \draw[fs/slot]  (135.8,17.6) rectangle (150.5,21.1);
  \draw[fs/arrow] (137,19.35) -- (149.3,19.35);
  \draw[fs/other] (127.3,13.2) rectangle (131.3,16.7);
  \draw[fs/slot]  (131.8,13.2) rectangle (142,16.7);
  \draw[fs/arrow] (133,14.95) -- (140.8,14.95);
  \draw[fs/arrow] (152.8,17.15) -- (157.5,17.15);
  \fill[fssharedtint] (165.5,15.9) rectangle (174.5,18.1);
  \draw[fs/time] (158.5,17) -- (176.6,17);
  \draw[fs/thin, draw=fsink] (165.5,15.9) -- (165.5,19);
  \draw[fs/release] (171,15.2) -- (171,18.8);
  \node[fs/label, anchor=south] at (166.5,19.1) {predicted time};
  \node[fs/label, anchor=north east] at (176.6,14.9) {randomized gate};
  \node[fs/label, anchor=north west] at (118.2,9.6) {Execution mode (\S\ref{sec:exec})};
  \node[fs/label, anchor=east] at (138.6,4.3) {original request};
  \draw[fs/hit] (139.3,2.8) rectangle (147.5,5.8);
  \draw[fs/ref] (147.5,4.3) -- (166.8,4.3);
  \node[fs/label, anchor=east] at (138.6,-1.4) {virtual request};
  \node[fs/slot, anchor=west, minimum width=26.5mm, minimum height=6.6mm, inner sep=0pt, align=center,
        text=fsink] (vr) at (139.3,-1.4) {with no reuse,\\filled by real requests};
  \draw[fs/arrow] (vr.east) -- (167,-1.4);
  \draw[fs/release] (167.3,-4.4) -- (167.3,6.3);
  \node[fs/label, anchor=west] at (168.1,0.95) {release};
  \end{scope}
\end{tikzpicture}
\caption{Overview of \sys. Left: a request with a hidden hit computes only the rest of its prompt, and the
release controller holds its response until its miss twin would have been ready. Right: the release
controller, enlarged. The routing rule chooses the miss twin's instance without seeing the hidden prefix.
Within that instance, either the prediction mode predicts the duration of each of the twin's steps from its
shape and releases at a randomized gate, or the execution mode runs a virtual request that does not reuse
the cache, filled by real requests where possible, and releases the original request when it finishes.}
\label{fig:overview}
\end{figure*}
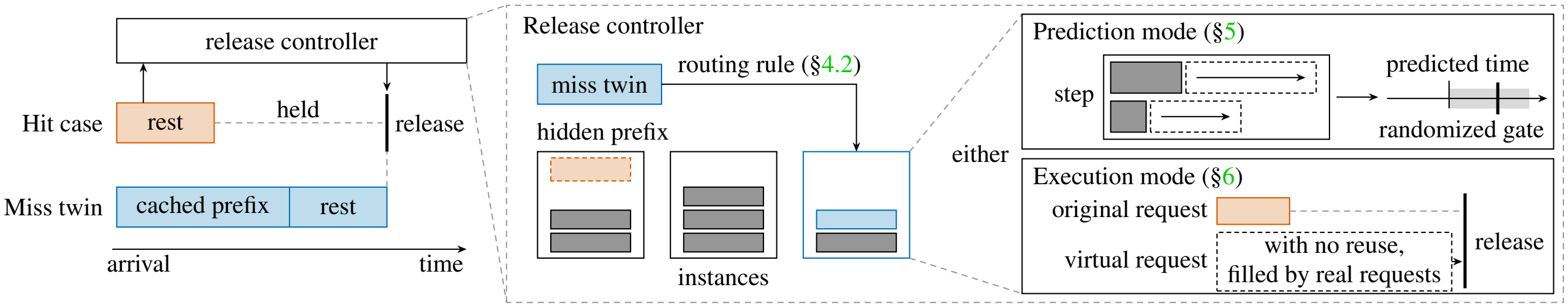

To hide such a hit, we let the request use the cache but delay its response: a \emph{release
controller} holds it until it would have been ready without the hit
(Figure~\ref{fig:overview}). That time depends on where the request would have run: cache-aware routing
favors the instance that holds the prefix~\cite{nvidia2026dynamorouter,llmd2026router,sglang2026gateway},
so without the hit, the request might have joined another instance's queue. It also depends on the prefill
work that the hit skips, which never runs and so cannot be timed. Published defenses each give up part of this goal. Some stop sharing, across
accounts~\cite{addagada2026governing,pang2024cache} or for content not deemed
public~\cite{chu2025selective,wu2026cacheprune}, and leave hits on what they still share visible. Others
let the first probe on each prefix through~\cite{pennas2026cachesolidarity,yan2026mirage}, or delay every
response, hits and misses alike~\cite{fonseca2026mitigacao} (Section~\ref{sec:priordesign}). None determines
where a request would have run without its hit.

Even if the request with the hit is delayed correctly, that may not be enough, because a hit can also affect
other requests: in laboratory settings, requests queued behind a hit can reveal it through the time it
saves them~\cite{prior2026savings}.
We define \emph{load visibility} as the extent to which clients can detect work done for another request,
prove that it bounds what such requests reveal, and give a test that any client can run. On three commercial
APIs, our requests could not tell another request's long prompt from a short one beyond measurement
noise (Section~\ref{sec:measure}). This suggests that their
clients cannot observe one another's load, so we target this setting, where a defense need only
protect each request's own timing. Our goal is to make each request's
release time indistinguishable from that of its \emph{miss twin}: the same request, allowed to reuse only
its own group's cache, while other requests' arrivals and hits stay fixed. Usage and billing also
follow the twin.

In a cluster, the twin's release time first depends on where the twin would run. The gateway
applies the deployment's routing rule to load and only to the cache entries visible to the request's
group, so a hidden hit cannot move the twin to another queue, while the request itself can still use the
hit on the instance that holds it (Section~\ref{sec:routing}). Because each request has its own release
target, requests that miss need little added delay (Section~\ref{sec:instance}).

Within the twin's instance, the release controller then obtains the twin's timing in one of two ways
(Figure~\ref{fig:overview}).
The \emph{prediction mode} predicts the duration of the twin's schedule under the current load and
releases the response through a randomized gate, following predictive
mitigation~\cite{askarov2010predictive,zhang2011predictive}.
The \emph{execution mode} instead runs that schedule: a \emph{virtual request} reserves the token
positions that the hit would skip, and waiting requests compute their own tokens there where possible.
The prediction mode suits deployments whose prefill durations are predictable, and the execution mode
those where the twin itself must set the release time.

We implement both modes on vLLM and evaluate them with every request isolated on up to 8 real
instances and 4,096 simulated instances. 
To examine their safety, we send probes and classify hits and misses by their timing. On a scale from 0 for guessing to 1
for certainty, the classifier's estimated advantage is at most 0.07 for both modes, against up to 0.98
under unprotected sharing (Table~\ref{tab:sec}). At saturation, the prediction mode retains 0.984 of unprotected throughput, more
than the 0.945 of isolation by account, and it adds at most 0.25\,s to the 90th-percentile time to first
token.

We make three contributions:
\begin{itemize}\setlength{\itemsep}{0pt}\setlength{\parsep}{0pt}\setlength{\topsep}{2pt}
\item Design principles for a prefill system that shares one cache yet isolates every request by default
(Section~\ref{sec:design}).
\item Measurements showing, to our knowledge for the first time, that unlike laboratory deployments,
commercial APIs do not let requests reveal one another's load, as captured by our load visibility
abstraction (Section~\ref{sec:threat}).
\item \sys, a vLLM implementation for clusters of prefill instances that makes hits and misses
indistinguishable in timing while any two requests can reuse each other's cache
(Sections~\ref{sec:routing} to~\ref{sec:exec}).
\end{itemize}

\section{System Model}\label{sec:sysmodel}

\noindent\textbf{Deployment.}
A cluster includes a gateway and $M$ prefill instances. Each instance runs a copy of the model on a group of
GPUs and has its own request queue, scheduler, and KV cache. The gateway routes each arriving request to an instance, favoring those that cache more of its
prefix~\cite{nvidia2026dynamorouter,llmd2026router,sglang2026gateway,srivatsa2025preble}, and streams the
resulting tokens back to the client.
With prefill-decode disaggregation, the prefill instance produces the first token and passes the KV cache
to a separate decode instance~\cite{zhong2024distserve,patel2024splitwise,qin2025mooncake}. Because a
cache hit saves prefill work but no decode steps, our model covers prefill through the first token.

\noindent\textbf{Steps.}
An instance computes prompts in \emph{steps}, each a forward pass over a batch of
\emph{chunks}, runs of consecutive tokens from one or more requests, as serving engines widely
do~\cite{kwon2023efficient,zheng2024sglang,agrawal2024taming}. All chunks in a step start
and finish together, so the step has one duration for all its requests, which our timing
predictions rely on. A step's token count is capped by a \emph{token budget}, so a prompt may span several steps.
A step's \emph{shape} comprises the number of requests, the token count of each chunk, and, for each chunk, the number of its tokens that were prefilled in prior steps.

\noindent\textbf{Prefix caching.}
Each instance caches the KV of prompt prefixes in fixed-size blocks~\cite{kwon2023efficient,zheng2024sglang}.
When the scheduler selects a request, the instance reuses the cached blocks at the start of its prompt
and computes only the remaining tokens. Each request belongs to a \emph{sharing group}, by default its own. A hit is \emph{visible} if the request's group has cached the matching blocks, and a hit on
blocks cached only outside the group is a \emph{hidden hit}, so with every request isolated, every hit is
hidden.

\noindent\textbf{Life of a request.}
The gateway routes request~$x$ to an instance, where it waits in the queue and then runs for one or more
steps. Steps are filled in first-come-first-served order, so no other work comes between $x$'s steps. Its time to first token is
\begin{equation}\label{eq:ttft}
T_x \;=\; g_x \;+\; \underbrace{r_x + \sum_{k \in W_x} d_k}_{\text{queueing}} \;+\;
\underbrace{\sum_{k \in E_x} d_k}_{\text{execution}},
\qquad d_k = d(\sigma_k).
\end{equation}
The \emph{gateway time}~$g_x$ includes network transit. It varies with load and path but is usually much
shorter than the time spent in steps, and it does not depend directly on whether $x$ hits.
The \emph{queueing time} includes the time~$r_x$ left in the current step when $x$ arrives and the
durations of the steps~$W_x$ that start before $x$ is selected. The \emph{execution time} sums the
durations of the steps~$E_x$ from the first to include $x$ through the step that produces its first token.
Because a step can also carry other requests, its duration~$d_k$ depends on the shape~$\sigma_k$ of the batch.

\section{What an Attacker Can Observe}\label{sec:threat}

\subsection{Two Kinds of Probes}\label{sec:probes}
A hidden hit can change the timing both of the request that uses it and of other requests, and an attacker
can observe either with a different kind of probe. A \emph{direct} probe starts with a guessed prefix and
observes its own time to first token, which arrives earlier when the probe reuses cached blocks (the
\emph{direct} channel)~\cite{song2025early,zheng2026inputsnatch,gu2025auditing}. A hit also changes its instance's
work, which affects every request served after it, hit or miss.
An \emph{indirect} probe therefore infers a hit from the timing of unrelated requests (the \emph{indirect}
channel)~\cite{wu2025know,wang2026optileak,prior2026savings}.

We evaluate both kinds of probes by their \emph{single-probe advantage}~$\eps$: given a hit or a
miss with equal probability, a classifier that sees one probe's timing guesses correctly with
probability at most $(1+\eps)/2$. This advantage measures cache membership inference, whether a
given prefix is cached, on which all cache side-channel attacks build, so bounding it also limits
attacks such as prompt extraction~\cite{zheng2026inputsnatch,wu2025know}. Membership matters even when
the attacker already knows the prefix. As with browser and DNS caches, whose contents are
public~\cite{felten2000timing,randall2020trufflehunter}, a hit on a published system prompt or a public
document reveals that some request used it and when, so an attacker can monitor a user's activity or
estimate how widely an application or agent harness is used. We therefore treat every prefix alike and
exempt no content as public. We consider one direct probe per prefix
because of \emph{cache pollution}: once the first probe has cached the prefix, later probes hit in both
cases until the blocks are evicted, and sending probes almost simultaneously does not avoid this (Supplement~\ref{app:evaldetail}).

\subsection{Load Visibility and Its Test}\label{sec:measure}
\noindent\textbf{Load visibility.}
On a large service, does sending a request change the response timing of an unrelated user's request by
an observable amount? Intuitively it does not, and a hit would disturb unrelated requests even less,
because a hit and a miss differ by less than sending the request and not sending it.
Unrelated requests do not use the probe's prefix, so they can observe a hit only if it
changes their \emph{load}:
the amount of work, the queue order, or the routing of requests. For a given
set of such requests, \emph{load visibility} is their
advantage in distinguishing another request's presence from its absence. Their advantage in distinguishing a hit
from a miss is at most the sum of the request's load visibilities in the two cases (Theorem~\ref{thm:lv}),
so where load is invisible, only direct probes need a defense. Load visibility can be measured
without a hit, even on a service that has yet to enable cache sharing.

\noindent\textbf{Test.}
Indirect probes nevertheless succeed in laboratory deployments~\cite{prior2026savings}, so a deployment
needs a test of whether they need a defense.
In each trial, a
request~A carries at random either a long fresh prompt or a 64-token prompt, and other 64-token requests
sent around it try to tell the two apart from their own first-token times. The long prompt adds more work
than the hits in scope would save, so it is easier to detect than a hit. A's own first-token time
serves as a control, since it must show the long prompt for the test to be valid.

We compare each resulting advantage bound with its noise bound, the level that measurement noise alone
reaches at that sample size (Section~\ref{sec:evalsetup}). The test passes for the routes tested if
every bound stays within noise, fails if a load effect is detected, and is otherwise inconclusive. When it
passes, twice the largest bound limits the advantage with which the tested requests distinguish a hit from
a miss, under the premise of Section~\ref{sec:worlds}.

\noindent\textbf{Commercial APIs.}
We ran the test on the Anthropic, OpenAI, and Gemini APIs, using \texttt{claude-haiku-5-5},
\texttt{gpt-5.6-luna}, and \texttt{gemini-3.5-flash-lite}, respectively, over one or more days.
The other requests were sent just before and after A, and Table~\ref{tab:measure} reads those sent after it. Every request used fresh random text and read no
cached tokens, and we measured only our own requests' latencies. Supplement~\ref{app:measure} gives further
routes, including requests that share a cached prefix with A so that cache-aware routing sends them
where it would send a hit, and we estimate advantages as in Section~\ref{sec:evalsetup}.

\begin{table}[t]
\centering\small\setlength{\tabcolsep}{5pt}
\caption{Load visibility on commercial APIs: advantage in distinguishing a long fresh prompt from a
64-token one, as an estimate and its 95\% upper bound, from the first-token time of the request
carrying the prompt (direct) and of one sent just after it (indirect).}
\label{tab:measure}
\begin{tabular*}{\columnwidth}{@{\extracolsep{\fill}}lrrrr@{}}
\toprule
 & \multicolumn{2}{c}{Direct probe} & \multicolumn{2}{c}{Indirect probe} \\
\cmidrule(lr){2-3}\cmidrule(l){4-5}
Service & Estimate & Upper bound & Estimate & Upper bound \\
\midrule
Anthropic & 0.997 & 0.999 & $-0.004$ & 0.019 \\
OpenAI    & 0.793 & 0.846 & $-0.010$ & 0.037 \\
Gemini    & 0.992 & 0.998 & $-0.010$ & 0.035 \\
\bottomrule
\end{tabular*}
\end{table}

\noindent\textbf{Results.}
Table~\ref{tab:measure} gives one run per service. Across all runs with prompts of at least 98K tokens,
A's own first-token time distinguished the long prompt with an estimated advantage of at least 0.76, so
the control holds. Other requests, by contrast, did
not: every indirect upper bound in Table~\ref{tab:measure} fell within its noise bound, 0.028 for Anthropic's run
and 0.055 to 0.072 for the other two, and the
largest, 0.037, limits such a request's advantage for a hit at 0.074 under that premise.
The other routes and statistics also stayed within noise, including in repeated runs
(Supplement~\ref{app:measure}).

\subsection{Threat Model and Goals}\label{sec:goals}
\noindent\textbf{Attacker.}
The attacker holds valid credentials for one or more accounts and chooses its prompts, including the
length of a guessed prefix, and when to send them. It observes the streamed tokens and usage fields of its
own requests but not the internal state of the gateway and instances, which are trusted, and it
influences routing only through the requests it sends. Every request is protected, whichever account the attacker uses.

\noindent\textbf{Security goal.}
For each request~$x$, the first-token release time at the gateway should have the same distribution as
its miss twin's (Section~\ref{sec:worlds}), and with prefill-decode disaggregation we extend this guarantee
to later tokens under the conditions of Section~\ref{sec:dnode}. Each mode bounds the single-probe
advantage by design, and Section~\ref{sec:evalsec} checks the design on the deployed code and with direct
probes, which capture every effect of a hit on $x$'s own timing, including a change of instance.
Table~\ref{tab:channels} gives the scope for other observations.

\noindent\textbf{Performance goal.}
Near full load, the cluster should approach the throughput of unprotected sharing, which lets any request
reuse any other request's cached KV, while keeping the deployment's existing cache-aware routing rule and
without offline work to fill spare capacity. Because added latency also matters, we also measure goodput under first-token SLOs
(Section~\ref{sec:evalsetup}).

\begin{table}[t]
\centering\footnotesize
\setlength{\tabcolsep}{3pt}
\caption{Channels through which a client can observe a hit and how \sys handles them.}
\label{tab:channels}
\begin{tabular}{@{}lll@{}}
\toprule
Channel & What a hit changes & \sys \\
\midrule
\multicolumn{3}{@{}l}{\emph{Timing}}\\
Own first token & arrives early & the twin's timing, \S\ref{sec:routing}, \S\ref{sec:pred}, \S\ref{sec:exec} \\
Later tokens & follow the first & only via load, \S\ref{sec:dnode} \\
Other requests & load, queue order & test must pass, \S\ref{sec:measure} \\
\addlinespace[3pt]
\multicolumn{3}{@{}l}{\emph{Returned values}}\\
Usage counts & cached tokens shown & the twin's count \\
Billing & cached tokens cheaper & billed as the twin \\
\addlinespace[3pt]
\multicolumn{3}{@{}l}{\emph{Content}}\\
Output numerics & may differ & out of scope \\
\bottomrule
\end{tabular}
\end{table}

\noindent\textbf{Side channels and scope.}
Beyond Table~\ref{tab:channels}, rate limits and other counters that discount cached tokens must count
hidden hits as misses. We consider deployments that pass the test of Section~\ref{sec:measure} for the
routes in scope. Decode on instances that also run prefill is outside our scope, as are physical
side effects.

\section{Design Principles}\label{sec:design}
The design has three steps: state what a release time must satisfy (Section~\ref{sec:worlds}), fix the
instance on which the miss twin would run (Section~\ref{sec:routing}), and choose the release target on that
instance (Section~\ref{sec:instance}).

\subsection{Security: Indistinguishable Release Times}\label{sec:worlds}
Let a request~$x$ hit a prefix cached outside its sharing group. %
Its \emph{miss twin} is the
same request, allowed to reuse only its own group's cache, with its arrival time and all other requests
unchanged, so the twin computes that prefix. We call serving $x$ with the hit the \emph{hit
case}~$\mathrm{h}$ and serving its twin the \emph{miss case}~$\mathrm{m}$. An attacker can produce both,
as probes with a right and a wrong guess of the same length, because a probe without a hidden hit is its
own twin. The defense knows which it serves, and the
attacker must not be able to tell.

For $w \in \{\mathrm{h}, \mathrm{m}\}$, let $Y_w$ be the \emph{ready time}, when the instance produces the
first token, and $Z_w$ the \emph{release time}, when the gateway releases it. A token cannot leave before it exists, so $Z_w \ge Y_w$. The attacker observes $Z_w$ and information common to both cases, and the gateway
time adds nothing to this, because it does not depend directly on the hit (Section~\ref{sec:sysmodel}).

\begin{definition}\label{def:secure}
Let $I$ be the information common to both cases.
A defense is \emph{$\eps$-secure} for $x$ if, given $I$, the total variation distance
$\mathrm{TV}(Z_{\mathrm{h}}, Z_{\mathrm{m}})$ is at most $\eps$.
\end{definition}

This distance is the largest single-probe advantage (Section~\ref{sec:probes}) that a classifier observing
one release time can attain.

\noindent\textbf{The indirect channel.}
Definition~\ref{def:secure} concerns only $x$'s own timing. For other requests, let case~$0$ be the hit case with $x$ removed, and let $O_w$ denote everything the attacker observes of other requests in
case~$w$. Because case~$0$ omits $x$, it is common to both cases, and the load visibility of $x$ in
case~$w$ is $\Lambda_w = \mathrm{TV}(O_w, O_0)$.

\begin{theorem}\label{thm:lv}
For any set of other requests,
$\mathrm{TV}(O_{\mathrm{h}}, O_{\mathrm{m}}) \le \Lambda_{\mathrm{h}} + \Lambda_{\mathrm{m}}$.
Under any coupling of each case with case~$0$ in which the release of $x$ is $\eps$-secure given $O_0$,
$\mathrm{TV}\bigl((Z_{\mathrm{h}}, O_{\mathrm{h}}), (Z_{\mathrm{m}}, O_{\mathrm{m}})\bigr)
\le \eps + \Pr[O_{\mathrm{h}} \ne O_0] + \Pr[O_{\mathrm{m}} \ne O_0]$.
\end{theorem}

The first bound follows from the triangle inequality through case~$0$ and holds for any queue or
scheduler. The second bounds the attacker's joint observation of the probe and other requests: it adds to
the direct bound~$\eps$ the probability, in each coupling, that $x$ changes any of those observations. Each such probability is at least $\Lambda_w$ for its case (Supplement~\ref{app:proofs}). To apply the
first bound empirically, Section~\ref{sec:measure} relies on one premise: an added request following $x$'s route and
requiring at least as much computation is at least as visible to the same requests. The empirical bound
then covers the routes and timing statistics tested.

\subsection{Routing: Reduce the Cluster to an Instance}\label{sec:routing}
Definition~\ref{def:secure} constrains two release times, and a ready time depends on the instance where
the request runs and on its queue. Estimating the twin's time accurately therefore first requires the
twin's instance, yet a cache-aware gateway can send $x$ to the prefix holder~$R$ in the hit case but to
another instance in the miss case.

\noindent\textbf{Routing the twin.}
The gateway therefore applies the deployment's routing rule at arrival to the load and the cache
entries visible to $x$'s sharing group. These inputs are the same in both cases, so they select the same
instance~$S$ for the twin, and with every request isolated, $S$ is the instance that the rule selects for a
request with no cache match.

\begin{proposition}\label{prop:reduce}
If all gateway decisions about $x$ use only inputs common to both cases, and the release of $x$ is
$\eps$-secure conditional on each outcome of those decisions, then its release is $\eps$-secure in the cluster.
\end{proposition}

The proof conditions on the gateway's decisions, which have the same distribution in both cases, and
therefore needs to consider only one instance at a time.

\noindent\textbf{Using a hit on another instance.}
The gateway can still send $x$ to $R$ to use the hit, and if a mode declines the hit, $x$ runs as a miss
on $S$ (Sections~\ref{sec:sharegate} and~\ref{sec:execcluster}). When $R \ne S$, choosing $R$ uses cache
entries outside the group and thus falls outside Proposition~\ref{prop:reduce}. The twin on $S$ still sets
the intended release time, however, so running $x$ on $R$ can change the release only if $x$ becomes ready
after that time. Hits with $R \ne S$ are common: in our security runs (Section~\ref{sec:evalsetup}), such requests
accounted for 0.34 to 0.37 of hit
tokens on two instances, 0.61 on four, and 0.78 on eight.

\subsection{Within Instance: One Target Distribution}\label{sec:instance}
\noindent\textbf{What the scheduler determines.}
Under first-come-first-served scheduling with a token budget, the queue at arrival determines $x$'s
waiting steps $W_x$, its execution steps $E_x$ (Equation~\eqref{eq:ttft}), and their shapes. The only exception is later arrivals that fill unused budget in the final steps, and near full load, they change the depths
of the added tokens but hardly their number. A deterministic scheduler can therefore reconstruct the
twin's steps even when its extra steps never run. Step durations, however, still vary with jitter at a
fixed shape, and the jitter of a step that never runs cannot be observed.

\noindent\textbf{One target distribution.}
Given $I$, let $F_{\mathrm{h}}$ and $F_{\mathrm{m}}$ be the distributions of $Y_{\mathrm{h}}$ and
$Y_{\mathrm{m}}$, and let $F^{-1}$ denote a quantile function. A release rule maps each ready time to
$Z_w = \phi_w(Y_w)$, where the map is nondecreasing and always satisfies $\phi_w(y) \ge y$; it can depend on
the case, because the defense knows whether $x$ hit. If both maps produce the same distribution~$Q$, the
defense is 0-secure for $x$.

\begin{theorem}\label{thm:target}
Possibly randomized maps with $\phi_w(y) \ge y$ can carry $F_{\mathrm{h}}$ and $F_{\mathrm{m}}$ to $Q$ if and only if
\begin{equation}\label{eq:dominance}
Q^{-1}(u) \;\ge\; \max\bigl\{F_{\mathrm{h}}^{-1}(u),\, F_{\mathrm{m}}^{-1}(u)\bigr\},
\quad u \in (0,1),
\end{equation}
that is, if $Q$ dominates both distributions at every quantile.
The right-hand side of~\eqref{eq:dominance} is the quantile function of a feasible target
$Q^{\ast}$ that every feasible target dominates.
\end{theorem}

Necessity follows because release cannot precede readiness~\cite{shaked2007stochastic}, and sufficiency
from the one-dimensional transport map $\phi_w = Q^{-1} \circ F_w$, which maps a ready time to the same
quantile of $Q$ and never releases early (Supplement~\ref{app:proofs}); with random tie-breaking, the
ready-time quantile is uniform. This monotone map minimizes the expected value of every convex cost of the
added delay~\cite{santambrogio2015optimal}, so it spreads the delay as evenly as possible.
For fixed execution, $Q^{\ast}$ thus minimizes every expected cost
that increases with release time. If $F_{\mathrm{m}}$ dominates $F_{\mathrm{h}}$, then
$Q^{\ast} = F_{\mathrm{m}}$ and misses need no delay. This dominance holds when a hit saves much more time
than a step's jitter, but it can fail when the savings are small or when the remaining chunks run more
slowly because they start deeper in the prompt.

\noindent\textbf{Cost.}\label{sec:cost}
A defense incurs a \emph{GPU cost} when it declines hits or computes tokens that serve no request, which we
measure in GPU time beyond unprotected sharing. Its \emph{latency cost} is the wait between readiness and
release. Holding a token uses no GPU time and leaves saturated throughput unchanged. Both costs, however, affect goodput under a time-to-first-token SLO: the GPU cost determines the throughput ceiling, and the latency
cost determines how loose the SLO must be to reach it.

\subsection{Why Existing Designs Fall Short}\label{sec:priordesign}
\begin{table}[t]
\centering\footnotesize
\setlength{\tabcolsep}{3pt}
\caption{Published designs, each judged within the isolation scope that it sets for itself.}
\label{tab:priordesign}
\begin{tabular}{@{}p{0.34\columnwidth}p{0.21\columnwidth}p{0.39\columnwidth}@{}}
\toprule
Design & \raggedright Secure in its scope & Shortfall \\
\midrule
\raggedright Sharing boundary fixed by account~\cite{addagada2026governing,pang2024cache} or
content~\cite{chu2025selective,wu2026cacheprune} & Yes &
\raggedright\arraybackslash Declines every hit across the boundary, at a GPU cost \\
\addlinespace[2pt]
\raggedright Boundary learned from reuse~\cite{pennas2026cachesolidarity} & \raggedright Except a prefix's first reuse &
\raggedright\arraybackslash The first probe decides membership (Definition~\ref{def:secure}) \\
\addlinespace[2pt]
\raggedright Delay to the miss time on the hit's instance~\cite{yan2026mirage} & \raggedright On one instance &
\raggedright\arraybackslash The instance depends on the hit (Proposition~\ref{prop:reduce}) \\
\addlinespace[2pt]
\raggedright Delay to a common target~\cite{fonseca2026mitigacao} & \raggedright Yes, also in a cluster &
\raggedright\arraybackslash Raises every request's time to first token (Theorem~\ref{thm:target}) \\
\bottomrule
\end{tabular}
\end{table}
Why do existing designs not suffice under these principles? We take each within the isolation scope that
it sets for itself, whether accounts, content, or prefixes, and consider its design rather than its
implementation (Table~\ref{tab:priordesign}).

\noindent\textbf{The first probe must be hidden.}
A design that fixes its sharing boundary in advance is secure within that boundary, at the GPU cost of
every hit that it declines (Section~\ref{sec:cost}). A design that learns the boundary from reuse instead
serves the first reuse of a prefix across users as an ordinary hit. Definition~\ref{def:secure} applies to
every request, and the first probe is the one that matters: a single probe decides whether the prefix is
cached, and it then caches the prefix itself, so an attacker gains nothing from probing again
(Section~\ref{sec:probes}). Such a design therefore protects only probes that an attacker has no need to
send.

\noindent\textbf{A hit's instance depends on the hit.}
A design that delays a hit to the miss time estimated on the instance where it runs hides the hit on one
instance. Under cache-aware routing, however, that instance is the prefix holder in the hit case and the
routing rule's choice in the miss case, so the two release times follow different queues and the premise
of Proposition~\ref{prop:reduce} fails (Section~\ref{sec:evalabl} measures this case).

\noindent\textbf{A common target raises every request's latency.}
A target common to all requests and instances, a fixed time or a coarse quantum, avoids this dependence
and is secure in a cluster as well. By Theorem~\ref{thm:target}, however, it must dominate the miss
distribution of every request on every instance under every load, so it raises the time to first token of
all requests, hits and misses alike. The minimal feasible target is instead the request's own, which
requires the twin's time (Sections~\ref{sec:pred} and~\ref{sec:exec}).

\section{Prediction Mode: Predicting the Twin}\label{sec:pred}

When both ready-time distributions are known and $F_{\mathrm{m}}$ dominates $F_{\mathrm{h}}$,
Section~\ref{sec:instance} gives an ideal release rule: instances execute as in unprotected sharing, misses
leave when ready, and each hit leaves at the corresponding quantile of its twin's ready time. For a hit,
however, $F_{\mathrm{m}}$ is unknown, because its twin never runs, and $F_{\mathrm{h}}$ is seen through a
single sample. Estimating both distributions and applying the one-dimensional transport of
Section~\ref{sec:instance} works poorly: any error in either estimate shifts hits against misses and
becomes leakage (Section~\ref{sec:gate}), each kind of hit offers few samples, and short hits can violate
dominance. The prediction mode therefore estimates a single quantity, the twin's ready time on $S$, and
builds a release rule around it that tolerates its error.

\subsection{Predicting the Twin's Steps}\label{sec:twinclock}

By Equation~\eqref{eq:ttft}, the twin's ready time is its queueing time plus the durations of its execution
steps, from the step in which it starts to the step that produces its first token. With a hit, these steps
never run as the twin would run them, but their shapes are known (Section~\ref{sec:instance}).
The twin starts in the first step after its arrival that has budget left
once the requests ahead of it are scheduled. Each of its steps then holds those requests' chunks at their
depths, the twin's own chunks, and tokens from requests that have since reached $S$, which the gateway
identifies from its routing log.

Given these shapes, the durations come from an offline linear model,
\begin{equation}\label{eq:steplaw}
d(\sigma) \;=\; a_0 + a_1 N + a_2 \sum_{i} n_i \Bigl(s_i + \frac{n_i}{2}\Bigr),
\end{equation}
where $N$ is the step's token count and chunk~$i$ has $n_i$ tokens starting at position $s_i$. With an online
correction factor per instance, its residual has a standard deviation of 2.2 to 2.4\,ms per step
(Supplement~\ref{app:rules}). Summing the model over the twin's steps gives the
twin's predicted ready time~$B$, which uses only requests routed to $S$. The residuals of recent misses
then estimate the distribution of miss ready times around~$B$. The residuals of hits, measured against predictions on the instances that serve them, instead inform cross-instance admission (Section~\ref{sec:sharegate}), and both estimates are kept separately for groups of similar requests (Supplement~\ref{app:rules}).

\subsection{Releasing Behind a Gate}\label{sec:gate}
\noindent\textbf{Sensitivity to prediction error.}
The transport map does not tolerate error in the prediction~$B$: with residual spread~$\varsigma$, a bias of magnitude~$\beta$ gives an advantage of about
$0.4\beta/\varsigma$, so at spreads of 10 to 30\,ms, keeping the advantage at or below 0.05 tolerates only 1
to 4\,ms of bias. In offline replay, this map indeed leaks more than the gate below (Table~\ref{tab:rules}). A robust rule must therefore merge or randomize ready times.

\noindent\textbf{A gate and its bound.}
The prediction mode releases the first token at $Z_w = \max(Y_w, V)$, where the gate~$V$ uses only inputs
common to both cases and an independent random draw. The gate's distribution thus serves as the target of
Section~\ref{sec:instance}. Because both cases use the same rule and a token ready by $V$ leaves at $V$,
coupling the two cases on the same gate gives
\begin{equation}\label{eq:main}
\eps \;\le\; \Pr\bigl[\,Y_{\mathrm{h}} \ne Y_{\mathrm{m}} \text{ and } \max(Y_{\mathrm{h}}, Y_{\mathrm{m}}) > V\,\bigr],
\end{equation}
since releases can differ only when ready times differ and at least one of them exceeds $V$. The bound
requires no distributional model, and for $k$ probes that do not affect one another's timing, the sum of
their bounds bounds their joint advantage.

\noindent\textbf{A randomized gate.}
We set $V = B + \gamma$. Let $F$ be the miss residual distribution and $c$ its $1-\eps$ quantile. For each request, we draw $\gamma$ between the median of $F$ and $c$ from the distribution function
$G(r) = \eps / (1 - F(r))$, which assigns mass $2\eps$ to the median and reaches 1 at $c$.

\begin{proposition}\label{prop:gate}
If $F$ is continuous and every hit is ready before $B$ plus the median of $F$, then the randomized
gate is $\eps$-secure. The release distributions agree strictly between the gate's lower endpoint and $B + c$.
They differ only in a mass~$\eps$ that hits place at the lower endpoint and misses place beyond $B + c$.
\end{proposition}

Proposition~\ref{prop:gate} sharpens the bound by using the shape of~$F$: for $\eps = 0.01$,
Inequality~\eqref{eq:main} alone gives about 0.05. A fixed gate at $B + c$ also leaks $\eps$, but under the
step jitter in our runs, the randomized gate holds requests 12 to 17\% less. When the residual distribution
is estimated, the two gates still leak equally, provided hits are ready before the gate's lower endpoint
and the true miss hazard rate stays at or below the estimated rate throughout the gate's range
(Supplement~\ref{app:proofs}).

\subsection{Hits That Finish Too Late}\label{sec:sharegate}
The gate's guarantee thus requires every hit to be ready before the gate's lower endpoint
(Proposition~\ref{prop:gate}); a hit that is ready later has only the weaker bound of
Inequality~\eqref{eq:main}. Two kinds of hits can be that late. First, a short hit saves little time. On
the instance selected for the twin, it is still used and held at the gate under the weaker bound, but the
gateway never moves a request to another instance to use a hidden prefix shorter than 1{,}024 tokens.
Second, a cross-instance hit runs on $R$ while its gate depends on the queue at $S$, and the two queues are
unrelated, so many such hits would be ready after the twin. The gateway therefore admits a cross-instance
hit only if it is predicted to finish before the twin, with margins from both residual distributions
(Supplement~\ref{app:rules}); otherwise, it declines the hit, and $x$ runs as a miss on $S$ as its own twin.

\section{Execution Mode: Virtual Requests}\label{sec:exec}

When $F_{\mathrm{m}}$ dominates $F_{\mathrm{h}}$, misses can leave when ready
(Theorem~\ref{thm:target}). To give a hit the same timing, the execution mode runs the hit's twin instead of
estimating it.

\begin{figure}[t]
\centering
\begin{tikzpicture}[fs/fig, x=1mm, y=1mm]
  \node[fs/panel] at (0,59) {a};
  \node[fs/label, anchor=north west] at (3,59) {Executing the twin};
  \node[fs/world] at (17,48.5) {Hit case};
  \node[fs/world] at (17,39.5) {Miss case};
  \foreach \k in {0,1,2} {
    \draw[fs/step] (18+\k*20.5,46) rectangle ++(19,5);
    \draw[fs/step] (18+\k*20.5,37) rectangle ++(19,5);
  }
  \draw[fs/other] (18.5,46.5) rectangle (24,50.5);
  \draw[fs/other] (18.5,37.5) rectangle (24,41.5);
  \draw[fs/slot]  (24.5,46.5) rectangle (36.5,50.5); \node[fs/label] at (30.5,48.5) {slot};
  \draw[fs/miss]  (24.5,37.5) rectangle (36.5,41.5); \node[fs/label] at (30.5,39.5) {$x$};
  \draw[fs/slot]  (39,46.5) rectangle (45.5,50.5);
  \draw[fs/hit]   (46,46.5) rectangle (57,50.5);     \node[fs/label] at (51.5,48.5) {$x$};
  \draw[fs/miss]  (39,37.5) rectangle (57,41.5);     \node[fs/label] at (48,39.5) {$x$};
  \draw[fs/hit]   (59.5,46.5) rectangle (68,50.5);   \node[fs/label] at (63.75,48.5) {$x$};
  \draw[fs/miss]  (59.5,37.5) rectangle (68,41.5);   \node[fs/label] at (63.75,39.5) {$x$};
  \draw[fs/other] (68.5,46.5) rectangle (77.5,50.5);
  \draw[fs/other] (68.5,37.5) rectangle (77.5,41.5);
  \draw[fs/ref] (78,35.5) -- (78,52.5);
  \node[fs/label, anchor=south] at (78,52.4) {$\tilde{Y}$};
  \node[fs/label, anchor=west] at (78.6,48.5) {$Y_{\mathrm{h}}$};
  \node[fs/label, anchor=west] at (78.6,39.5) {$Y_{\mathrm{m}}$};
  \draw[fs/time] (18,33) -- (84,33);
  \node[fs/label, anchor=north east] at (84,32.7) {time};
  \node[fs/panel] at (0,29) {b};
  \node[fs/label, anchor=north west] at (3,29) {Filling the slot};
  \draw[fs/ref] (32,7) -- (32,25.5);
  \draw[fs/ref] (70,7) -- (70,25.5);
  \node[fs/world] at (17,22.5) {Slot};
  \node[fs/world] at (17,16.5) {Longer $x'$};
  \node[fs/world] at (17,10.5) {Shorter $x'$};
  \draw[fs/slot]   (32,20.5) rectangle (70,24.5); \node[fs/label] at (51,22.5) {$n$ tokens};
  \draw[fs/cached] (18,14.5) rectangle (32,18.5);
  \draw[fs/recomp] (32,14.5) rectangle (40,18.5);
  \draw[fs/req]    (40,14.5) rectangle (70,18.5); \node[fs/label] at (55,16.5) {$x'$};
  \draw[fs/later]  (70,14.5) rectangle (80,18.5);
  \draw[fs/cached] (18,8.5) rectangle (32,12.5);
  \draw[fs/req]    (32,8.5) rectangle (52,12.5);  \node[fs/label] at (42,10.5) {$x'$};
  \draw[fs/fab]    (52,8.5) rectangle (70,12.5);
  \draw[fs/time] (18,7) -- (84,7);
  \node[fs/label, anchor=north east] at (84,6.7) {depth};
  \foreach \x/\t in {18/$0$, 32/$s$, 40/$p$, 52/$s+\ell$, 70/$s+n$} {
    \draw[fs/thin] (\x,7) -- (\x,6.2);
    \node[fs/label, anchor=north] at (\x,6.1) {\t};
  }
  \draw[fs/cached] (18,0) rectangle ++(4,2.6);  \node[fs/label, anchor=west] at (22.3,1.3) {read};
  \draw[fs/recomp] (32,0) rectangle ++(4,2.6);  \node[fs/label, anchor=west] at (36.3,1.3) {recomputed};
  \draw[fs/fab]    (52,0) rectangle ++(4,2.6);  \node[fs/label, anchor=west] at (56.3,1.3) {dummy};
  \draw[fs/later]  (71,0) rectangle ++(4,2.6);  \node[fs/label, anchor=west] at (75.3,1.3) {later};
\end{tikzpicture}
\caption{(a) Each frame is a full step, widths indicate token counts,
and gray boxes denote other requests. A virtual request reserves the $n$ positions that $x$ skips on a
hidden hit, so the first token of $x$ is ready at the end of the same step in both cases. (b) The sharing
group of $x$ has cached $s$ tokens, and a waiting substitute $x'$ fills the slot from depth $s$ to $s+n$
with its own work. A substitute with $p>s$ cached tokens recomputes $p-s$ of them, and if only $\ell<n$ of
its tokens remain, dummy tokens fill the last $n-\ell$ positions.}
\label{fig:exectwin}
\end{figure}
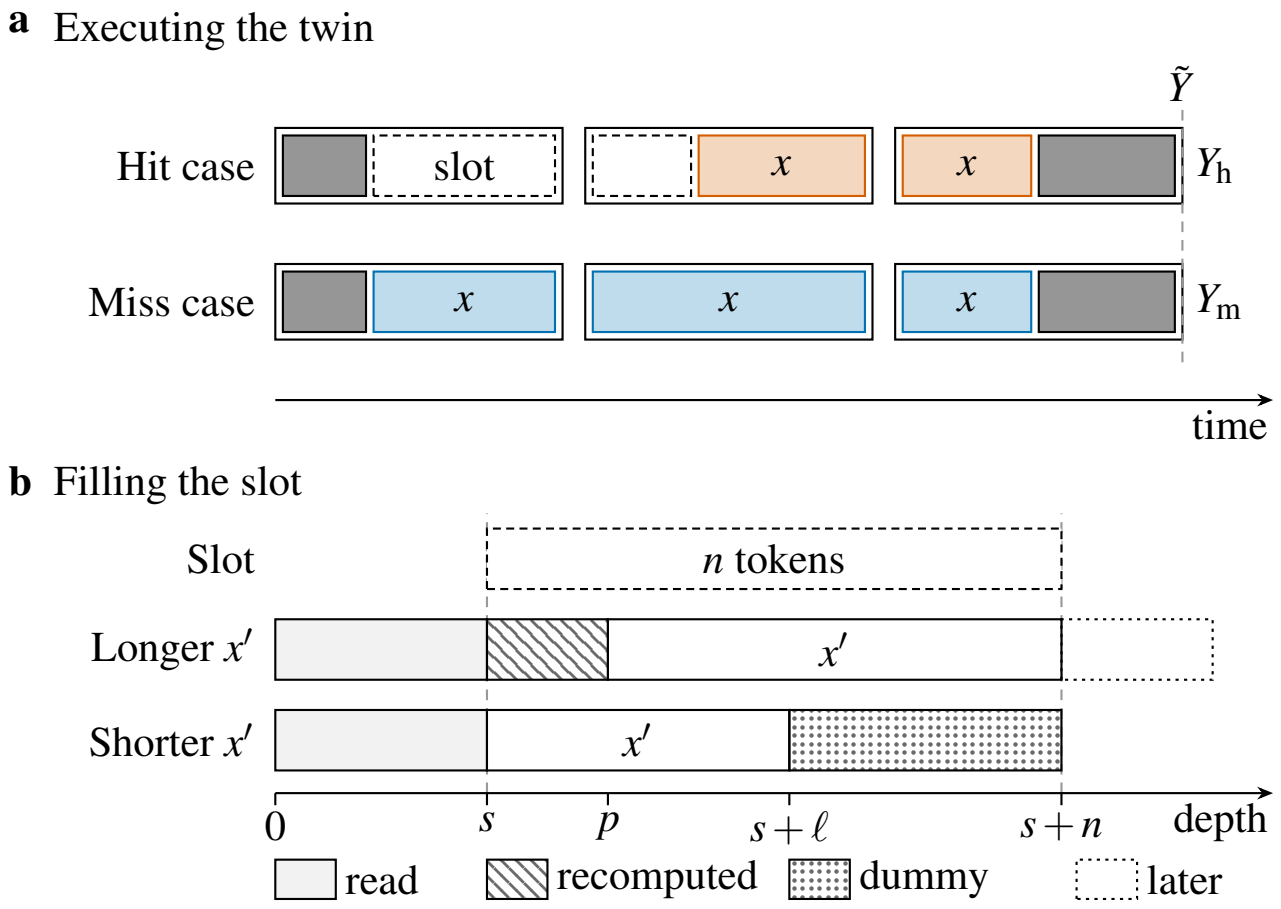

\subsection{Executing the Twin}\label{sec:vreq}
Suppose $x$ arrives with a hidden hit of $n$ tokens starting at depth~$s$, where $s$ is the part of the
prefix that its own sharing group has cached, zero when every request is isolated. The instance~$S$
selected for the twin receives a \emph{virtual request}, which stands for the work that the twin performs
and $x$ skips. The request $x$ itself still uses the hit, but its first token is not released before the
virtual request has finished, so $x$ leaves at its twin's time.

When $x$ runs on $S$, the virtual request takes $x$'s place in the queue and reserves the $n$ positions
from depth $s$ to $s+n$, its \emph{slot}, and $x$ follows it (Figure~\ref{fig:exectwin}a). The scheduler
treats the slot like any request, so in each step the slot and $x$ together occupy the same budget and
positions as the miss twin, and the chunks of $x$ are the twin's final chunks. Hence, if $\tilde{Y}$ is the
end of the hit-case step that holds the twin's last chunk, then $Y_{\mathrm{h}} = \tilde{Y}$ in every run
in which the slot runs as designed. When $x$ instead runs on the holder $R \ne S$, the gateway sends it to
the head of the queue on $R$ so that it finishes early, and the slot on $S$ covers the twin's entire
computation, whose final step determines $\tilde{Y}$.

A step with fewer tokens would finish sooner, so in every step that contains part of the slot, another
request computes its own tokens in those positions (Section~\ref{sec:fill}). Each step thus keeps the twin's token counts and depths. The step where the slot ends holds one more
request, since the slot and $x$ count as two, a deviation that Section~\ref{sec:execsec} bounds. The hit therefore does not make
$x$ faster; the computation it saves goes to whatever fills the slot. Matching step shapes reproduces
timing only when duration depends on shape alone, and Section~\ref{sec:content} treats content-dependent
durations.

\subsection{Filling the Slot}\label{sec:fill}
The slot must hold computation of the same shape, and useful computation is better than none. A
\emph{substitute} is a waiting request~$x'$ that does useful work in the slot while preserving its shape
(Figure~\ref{fig:exectwin}b). To match the starting depth, the sharing group of $x'$ must have cached at
least its first $s$ tokens; the substitute reads exactly those and recomputes any cached tokens it meets
in the slot, because a chunk that starts at a different depth has a different cost. To fill the reserved
length, a substitute that extends beyond the slot computes the rest later as an ordinary request, and
\emph{dummy tokens} fill whatever a shorter one leaves, or the entire slot if no request qualifies. The
instance keeps any block that contains a dummy token out of the prefix cache.
A substitute must also be unstarted and wait at least one full token budget of queued work behind $x$, so
that the requests that would fill the twin's final step in the miss case stay in place. Among eligible
requests, the scheduler chooses the one that wastes the least computation.

\noindent\textbf{Substitutes from the gateway.}
A substitute on one instance must already wait behind $x$, and a short queue often has none. The arrivals
of the whole cluster offer far more candidates, and the denser they are, the more likely one fits a slot
with little waste, so the gateway also reserves substitutes. It holds every request for a minimum dispatch delay, the same for all
requests, which costs only latency. When it dispatches $x$ with a hidden hit, it can reserve a request
that is still waiting out this delay, provided that the request has no block of its own group cached
anywhere and is chosen using only information common to both cases. The instance sets a reserved request aside until the
slot takes it or its own dispatch time arrives, when it joins the queue as it would have without the
reservation. The gateway decides at dispatch from its own state, and its router counts a hit's pending
prefill work as its twin's, so that the load it sees does not depend on the hit. In light-load offline
replays, the pool of candidates grew from 0.8 requests per instance with two instances to 13 with 32.

Recomputed prefixes, dummy tokens, and declined cross-instance hits (Section~\ref{sec:execsec}) make up
this mode's GPU cost, and the minimum dispatch delay and holding hits to their twin's time make up its
latency cost (Section~\ref{sec:cost}).

\subsection{Security}\label{sec:execsec}
A hit is released at $Z_{\mathrm{h}} = \max(Y_{\mathrm{h}}, \tilde{Y})$, whereas on a miss, $x$ is its own
twin and leaves at $Z_{\mathrm{m}} = Y_{\mathrm{m}}$.

\begin{proposition}\label{prop:exec}
Given $I$, the execution mode is $\eps$-secure with
\begin{equation}\label{eq:exec}
\eps \;\le\; \Pr\bigl[\,Y_{\mathrm{h}} > \tilde{Y}\,\bigr] \;+\; \mathrm{TV}\bigl(\tilde{Y},\, Y_{\mathrm{m}}\bigr).
\end{equation}
\end{proposition}

\noindent
The hit's release time can differ from $\tilde{Y}$ only if $Y_{\mathrm{h}} > \tilde{Y}$, so the probability
of that event bounds $\mathrm{TV}(Z_{\mathrm{h}},\tilde{Y})$, and the triangle inequality then
gives~\eqref{eq:exec}.

\noindent\textbf{On one instance.}
The first term asks more than Theorem~\ref{thm:target}, because execution pairs each hit with its own twin
rather than with any feasible target quantile, but it is zero whenever the slot runs as designed, since
then $Y_{\mathrm{h}} = \tilde{Y}$.
Details of the implementation that shift $\tilde{Y}$ by $\Delta$ add to the second term at most about
$0.4\,|\Delta|/\varsigma$ for normally distributed timing with standard deviation~$\varsigma$
(Supplement~\ref{app:proofs}), and a whole step that changes with probability~$\theta$ adds at most~$\theta$.
Section~\ref{sec:evalsec} checks on the deployed scheduler code whether the slot runs as designed.

\noindent\textbf{Across instances.}\label{sec:execcluster}
When $R \ne S$, $x$ and its twin face different queues, so the first term of~\eqref{eq:exec} is no longer
zero by construction. The gateway therefore admits a cross-instance hit only if $x$'s predicted ready time on $R$ plus 500\,ms precedes its twin's predicted ready time on $S$, and otherwise runs $x$ as a miss on $S$. The prediction thus decides only admission, while
the release still follows the executed twin, and in our two-instance security runs only 39 of 24{,}108
admitted hits became ready after their twins.

\noindent\textbf{Bait requests.}
Substitution also changes when other requests are served: a substitute computes in the slot ahead of
requests that arrived before it, so steps are no longer filled strictly first come, first served
(Section~\ref{sec:sysmodel}). Whether a request is drawn depends on a hidden hit, an effect on other
requests that Theorem~\ref{thm:lv} covers. An attacker can aim at it with a \emph{bait}: just after a
probe, it sends a request that fits the probe's slot and infers a hit if the bait's first token arrives
early. An early bait, however, need not be due to the probe. Any request's hidden hit creates a slot that
can draw the bait, and with every request isolated by default such hits are common. Nor can the attacker
make the probe's slot take the bait, because the gateway chooses among all requests in the cluster that
are waiting out the dispatch delay. When we ran this attack, the bait's timing gave an estimated advantage
of at most 0.027 (Supplement~\ref{app:cdt}, Table~\ref{tab:bait}).

\section{Evaluation}\label{sec:eval}

\subsection{Implementation and Setup}\label{sec:evalsetup}
\noindent\textbf{Implementation.}\label{sec:impl}
\sys runs on vLLM~0.26~\cite{vllm2026v026source}: its gateway is about 1{,}600 lines of Python in
one asynchronous process, and vLLM loads another 4{,}500 lines as a custom scheduler class. The gateway
tracks each instance's cache exactly, both overall and by sharing group, and labels each request with its
group, which is the request alone unless noted. We used AI coding agents to help implement \sys.

\noindent\textbf{Testbed.}
Each instance runs Llama-3.1-8B-Instruct on one NVIDIA L40S GPU with 16-token cache blocks, a 2{,}048-token
step budget, and a 147{,}200-token KV pool. Each request ends at its first token, as on a prefill instance.

\noindent\textbf{Workloads.}
We use two synthetic workloads and one from a public trace. In the \emph{self-reuse workload}, requests
mostly reuse their own account's cache, as when they continue a conversation, so isolation by account
keeps most of the reuse. The \emph{cross-reuse workload} shifts reuse to documents shared across accounts,
which isolation by account forgoes. The third comes from the public Mooncake conversation
trace~\cite{qin2025mooncake}. With every request isolated, all reuse in these workloads is hidden. Runs on
four and eight instances interleave copies of a workload, and Supplement~\ref{app:evaldetail} gives each
workload's composition and cached share.

\noindent\textbf{Configurations.}
We compare five configurations that serve the same requests with the same cache-aware routing rule. The
three baselines are unprotected sharing, isolation by account, which allows reuse only within an account,
and a \emph{no-reuse} baseline, in which every request computes its whole prompt. \sys runs in the
prediction mode, with the randomized gate at $\eps = 0.01$, and in the execution mode.

\noindent\textbf{Load and goodput.}
A closed-loop client keeps a fixed number of requests in flight at each of nine stages, which span short
to long queues. We use it for the security tests and to measure each configuration's saturated throughput
relative to unprotected sharing. An open-loop client offers Poisson arrivals up to 1.04 times the
saturated throughput of unprotected sharing, or of the configuration itself for the execution mode and the
no-reuse baseline (Supplement~\ref{app:evaldetail}). At a given time-to-first-token SLO, goodput is the
highest arrival rate at which 90\% of requests meet the SLO at the client, normalized to unprotected
sharing's saturated throughput.

\noindent\textbf{Probe trials.}
Security runs interleave one probe trial per four other requests, and each trial samples the hit or the
miss case of Section~\ref{sec:worlds} with equal probability. A victim on a fresh account sends a prefix of
2{,}048, 7{,}168, or 16{,}384 tokens followed by a fresh tail, prefix $P$ in the hit case or a different
prefix in the miss case. The probe, from another fresh account unless noted, then sends $P$ with a fresh tail 0.2\,s after the
victim's first token. Probes account for about half of the computation in these runs, so we report
throughput from runs without probes unless noted.

\noindent\textbf{Advantage and noise bound.}
We select a threshold on first-token time using half of the probes and, on the held-out half, estimate the
classifier's single-probe advantage and its one-sided 95\% bootstrap upper bound
(Supplement~\ref{app:tvthr}). The \emph{noise bound} is the simulated 95th percentile of this upper bound
without any effect, 0.088 at 1{,}000 probes per case and 0.029 at 9{,}000. Without any effect, a cell's
upper bound exceeds its noise bound with probability 0.05, so an upper bound above it is evidence of a hit,
whereas one at or below it shows only that this probe found none.

\begin{figure*}[t]
\centering
\includegraphics{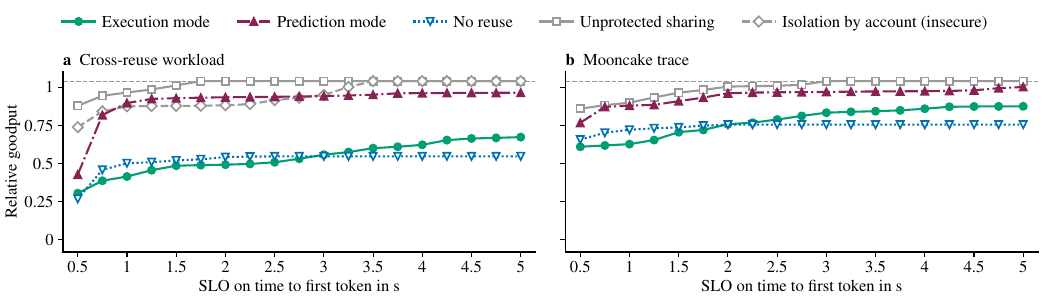}
\caption{Open-loop goodput on eight instances, relative to unprotected sharing's saturated throughput.}
\label{fig:goodput}
\end{figure*}
\subsection{Security}\label{sec:evalsec}
\begin{table}[t]
\centering\small
\caption{Single-probe advantage of the timing classifier.}
\label{tab:sec}
\begin{tabular*}{\columnwidth}{@{\extracolsep{\fill}}lrrrr@{}}
\toprule
Arm & Hit & $M{=}2$ & $M{=}4$ & $M{=}8$ \\
\midrule
Unprotected       & 7168  & 0.210 & 0.184 & 0.170 \\
                  & 16384 & 0.980 & 0.937 & 0.929 \\
Prediction        & 7168  & 0.000 & 0.002 & $-0.006$ \\
                  & 16384 & $-0.016$ & $-0.004$ & 0.068 \\
Execution         & 7168  & $-0.001$ & 0.003 & $-0.006$ \\
                  & 16384 & 0.006 & $-0.030$ & $-0.020$ \\
\bottomrule
\end{tabular*}
\end{table}

We ask whether direct probes detect hidden hits under either mode of \sys.

\noindent\textbf{Measured advantage.}
Table~\ref{tab:sec} gives the threshold classifier's advantage in each cell. Under unprotected sharing,
the probes are sensitive to hits of both lengths at every cluster size and almost always detect a
16{,}384-token hit. A 2{,}048-token hit saves only 20 to 40\,ms at the median against queueing delays of
seconds, so pooled over load stages the classifier reads nothing even under unprotected sharing, and we
list those cells only in Table~\ref{tab:secfull}. With \sys, every upper bound lies at or below its noise
bound, which is 0.03 to 0.06 for 7{,}168-token hits and, with fewer probes, 0.18 to 0.23 for 16{,}384-token
hits. Read in finer cuts, by groups of stages and, for the execution
mode, among probes whose twin ran on the prefix holder, the same probes give 62 upper bounds when the
eighteen pooled cells are counted. Of these, 3 exceed their noise bounds,
where about three would be expected without any effect (Supplements~\ref{app:cdt}
and~\ref{app:evaldetail}).

\begin{table}[t]
\centering\small
\caption{Consistency check of the execution mode's scheduler: scenarios run with and without $x$'s hit,
and those in which $x$'s first token differed, each time later in the hit case.}
\label{tab:cdtsum}
\begin{tabular*}{\columnwidth}{@{\extracolsep{\fill}}lrr@{}}
\toprule
Slots filled by & Scenarios & Differ \\
\midrule
Dummy tokens only & 12{,}673 & 0 \\
Substitutes, as deployed & 9{,}673 & 28 \\
\quad reserved request tied with $x$ at dispatch & & 21 \\
\quad $x$ arrived during the victim's prefill & & 3 \\
\quad $x$ itself drawn as a substitute & & 4 \\
\bottomrule
\end{tabular*}
\end{table}

\noindent\textbf{Consistency with and without a hit.}\label{sec:evalcdt}
For each mode, we also test the property on which its analysis rests. The execution mode's guarantee rests on its scheduler taking the
same steps for $x$ in both cases (Section~\ref{sec:execsec}), so our main evidence for this mode is a check
of that property on the deployed code. A harness drives the scheduler class and the gateway's reservation
logic through the same arrivals on one instance twice, with and without $x$'s hit, under step times that
depend only on shape, and compares every request's first-token step and usage fields
(Table~\ref{tab:cdtsum}, Supplement~\ref{app:cdt}).

With slots filled only by dummy tokens, the two runs agree for every request. With substitutes as
deployed, other requests differ by design, because the saved work goes to them, and $x$'s own first token
differs in 28 scenarios, with the three causes of Table~\ref{tab:cdtsum}. A reserved request tied with
$x$ at dispatch only on the regular arrival grid of the enumerated scenarios and in none of the real runs
whose logs we examined, and our probes do not arrive during the victim's prefill. In the remaining four,
$x$ was itself drawn as a substitute for another slot, and its first token came one step later. All 28
differences involve substitutes. Section~\ref{sec:execsec} analyzes this effect and reports the result of
the bait attack that targets it.

\begin{table}[t]
\centering\small
\setlength{\tabcolsep}{3.5pt}
\caption{Release rules on the prediction mode's security runs, with about 7{,}650 probes per case. Miss
wait is the mean delay added to misses, and for each hit length in tokens, entries give the estimate and
upper bound of the single-probe advantage when $B$ is known. The first three rows report the largest values
over three random draws per rule in replay, and the last row measures the deployed gate.}
\label{tab:rules}
\begin{tabular}{@{}lccc@{}}
\toprule
Rule & Miss wait & 7168 & 2048 \\
\midrule
Transport, hits only         & 0\,ms       & 0.057 / 0.071 & 0.013 / 0.032 \\
Quantile band, $w = 0.2$     & 1.0\,ms     & 0.047 / 0.058 & 0.039 / 0.052 \\
Gate, $\eps = 0.01$ & 20.4\,ms    & 0.018 / 0.032 & 0.008 / 0.030 \\
Gate, deployed               & 15.3\,ms    & 0.004 / 0.023 & 0.005 / 0.025 \\
\bottomrule
\end{tabular}
\end{table}
\noindent\textbf{A stronger observer.}
For the prediction mode, we test release rules against a stronger observer, who also knows each request's
predicted twin time $B$, which no client can see (Table~\ref{tab:rules}). Even then, the deployed gate
gives the smallest upper bounds for both hit lengths while delaying misses by 15\,ms on average. For
7{,}168-token hits, delaying only hits with the transport map of Section~\ref{sec:instance} gives the
largest, and a \emph{quantile band}, which randomizes each request's quantile within a band of
width~$w$, reduces it only partly (Supplement~\ref{app:rules}).

\noindent\textbf{Routing ablation.}\label{sec:evalabl}
Routing the twin is necessary: with the twin routed instead to the instance where the request runs, the
prediction mode was detected as an instance joined routing, with an upper bound of 0.433 against a noise
bound of 0.116, and with the twin at $S$ it was not, at 0.074 (Supplement~\ref{app:prior}).

In summary,
direct probes found no hit in any pooled cell of either
mode.

\subsection{Throughput and Latency}\label{sec:evalperf}

\noindent\textbf{Goodput under an SLO.}
Figure~\ref{fig:goodput} shows open-loop goodput on eight instances without probes. Holding costs the
prediction mode goodput mainly at the tightest SLO: from 0.75 to 1.5\,s it already delivers 0.87 to 0.99
of unprotected sharing's goodput, and on the cross-reuse workload it leads isolation by account from 1\,s
on. The execution mode delivers 0.48 and 0.73 of unprotected sharing's goodput on the cross-reuse workload
and the Mooncake trace at an SLO of 1.5\,s.

\noindent\textbf{Saturated throughput.}
In the closed loop on two instances without probes, the prediction mode retains 0.984 of unprotected
throughput on the self-reuse workload, 0.977 on the cross-reuse workload, and 0.984 on the Mooncake trace,
against 0.945 and 0.802 for isolation by account on the two workloads with account labels
(Supplement~\ref{app:perf}). Per completed prompt token, it realizes 99\% of the GPU-time savings that
unprotected sharing has over the no-reuse baseline. With every request isolated, the execution mode delivers 1.07 to 1.26 times the throughput of the
no-reuse baseline on the three workloads but trails isolation by account on both workloads with account labels. The scheduler often finds
no waiting request of the right shape for a slot: of all tokens computed on the self-reuse workload, 10\%
do substitute work and 39\% are dummies, including the slots of cross-instance twins. Longer queues offer
more matching requests (Section~\ref{sec:execcost}), and Figure~\ref{fig:scale} estimates larger scales.

\noindent\textbf{Which hits must be hidden.}
The execution mode's cost also depends on which hits it must hide. On the self-reuse workload, 89\% of
reusable tokens are reused within a conversation, so in one run on two real instances without probes, we
made each conversation a sharing group. Dummy tokens then accounted for only 4.0\% of the tokens the
execution mode computed, and its throughput was 1.015 times that of unprotected sharing
(Supplement~\ref{app:execload}).

\begin{figure}[t]
\centering
\includegraphics{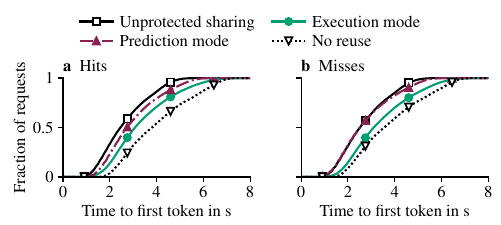}
\caption{Time to first token on eight instances.}
\label{fig:cdf}
\end{figure}
\noindent\textbf{Latency.}
On eight instances, holding under the prediction mode raises the median time to first token
of hits by 0.28\,s over unprotected sharing and leaves that of misses unchanged (Figure~\ref{fig:cdf}). Without probes on the
self-reuse workload, the prediction mode holds hits for 340\,ms and misses for 19\,ms on average, and
across stages and workloads it adds at most 0.25\,s to the 90th-percentile time to first token. The
execution mode and the no-reuse baseline, whose extra computation lengthens their queues, add more than
1\,s on the self-reuse workload (Supplement~\ref{app:perf}).

\section{Discussion}\label{sec:discussion}

\subsection{Larger Clusters and the Two Modes}\label{sec:evalscale}\label{sec:execcost}
Production clusters can have far more instances than our eight. A larger cluster serves more of the
request population and holds more KV cache, so a request more often finds its prefix already computed, and
the execution mode can draw substitutes from the arrivals of more instances (Section~\ref{sec:fill}).

\noindent\textbf{How much reuse is available?}
The \emph{reusable share} $H$ is the fraction of prompt tokens
that an earlier request in the cluster has already computed. It is an upper bound on the cached share,
because finite GPU memory and routing can both prevent reuse.
At saturation, throughput per instance relative to the no-reuse baseline equals the number of prompt tokens
served per token computed. If all available hits are used, unprotected sharing and the prediction mode can therefore reach $1/(1-H)$. The execution mode, however, is bounded by $1+H$, less than twice the throughput of
the no-reuse baseline, because a substitute computes its whole prompt and has hits only as often as other
requests.

\noindent\textbf{Estimating larger clusters.}
Our testbed supports up to eight real instances, or 32 emulated ones that replace the forward pass with a
delay from a measured step-time model. To estimate throughput beyond it, we vary
\emph{coverage}: a cluster of $M$ instances serves a random fraction $M/4{,}096$ of a fixed population drawn
like the self-reuse workload. An event-level simulator runs 16 to 256 instances on proportionally smaller
populations, and we calibrate throughput models on these runs; Supplement~\ref{app:scale} validates
both against real and emulated runs.

\begin{figure}[t]
\centering
\includegraphics{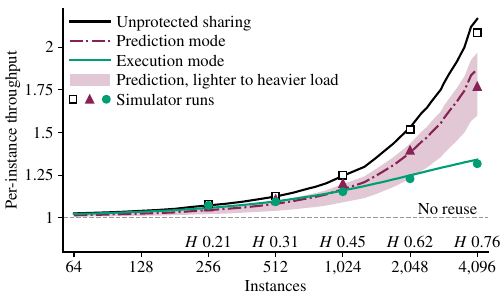}
\caption{Estimated throughput per instance relative to the no-reuse baseline, for a cluster of $M$
instances; $H$ is the reusable share of
prompt tokens.
}
\label{fig:scale}
\end{figure}

\noindent\textbf{The prediction mode at high reuse.}
At the load used for calibration, 14.17 requests in flight per instance, the two modes' normalized throughputs differ by at most 0.02 up to about $H=0.4$,
and above that the prediction mode leads (Figure~\ref{fig:scale}). At full coverage, $H=0.758$, it delivers
1.77 to 1.87 times the throughput of the no-reuse baseline, against 1.32 to 1.34 for the execution mode,
with the low ends from the simulator. The prediction mode then retains 0.85 to 0.86 of unprotected throughput. Routing explains most of its gap to the ideal of 4.13: even unprotected sharing reaches only
2.08 to 2.17, because the router often chooses a less loaded instance or one that holds a shorter prefix.

\noindent\textbf{What limits the execution mode.}
Dummy tokens keep the execution mode below its ceiling of 1.76 at full coverage: in emulation, they made
up 0.26 to 0.39 of the tokens computed. This waste shrinks with longer queues, so load matters more than
size. On eight instances, one real and seven emulated, raising the load from 9.2 to 12.5 requests in
flight per instance raised the share of the throughput gap between the no-reuse baseline and unprotected
sharing that the execution mode closed from 0.11 to 0.23, whereas averaged over these loads that share
stayed between 0.17 and 0.18 on 2 to 32 instances.

The two modes are thus two design points. At heavier load, prediction errors widen the margin for admitting
cross-instance hits, so the prediction mode gives up more reuse: at the heavy-load edge of the band in
Figure~\ref{fig:scale}, it overtakes the execution mode only above about $H=0.65$. The execution mode, in
contrast, times release by the twin itself and uses its slots better when queues are long.

\subsection{Content-Dependent Steps}\label{sec:content}
Both modes match timing through step shape. In a mixture-of-experts model,
however, timing also depends on which experts the tokens activate, and on Qwen3-30B-A3B at light load, the
execution mode without a fix exposes hits with single-probe advantages of 0.71 to 0.87.
The fixes therefore model content as well. In the execution mode, the dummy tokens that fill a slot repeat
the hit's own prefix, so they activate the same experts as the twin. The prediction mode uses the slowest
step times observed for the prompt's kind of content. Across three kinds of prefix, natural text, code,
and a repeated token, and three loads, the
single-probe advantages were then at most 0.034 for the execution mode and 0.023 for the prediction mode,
each in its own runs. In open-loop runs on the Mooncake trace, the two modes then delivered 0.66 to 0.84
and 0.88 to 0.99 of unprotected sharing's goodput at SLOs of 0.5 to 5\,s, against 0.62 to 0.71 for the
no-reuse baseline (Supplement~\ref{app:ext}). Dense models show smaller content effects, which these probes do not target.

\subsection{Multi-GPU Instances}\label{sec:tp}
An instance can itself span several GPUs. Under tensor parallelism, each layer ends in an all-reduce
whose duration could vary even at a fixed step shape. On two Qwen3-32B instances, each spanning two
NVLink-connected A100s, we recalibrated the prediction mode's step-time model, whereas the execution mode,
which times release by its twin, kept its L40S timing parameters. For 7{,}168-token hits, both modes' upper bounds on the single-probe
advantage were 0.043, below the noise bound of 0.069 (Supplement~\ref{app:ext}).

\subsection{Disaggregated Decode}\label{sec:dnode}
Under disaggregation, later tokens come from a decode instance. In \sys, that instance receives $x$ when its first token is released, fetches its KV cache
at once, admits it after a delay determined only by the prompt length, and caches no prefixes. The
advantage from observing all token times is then at most the advantage on the first token plus the mean
distance between the decode instance's other inputs, its other arrivals and its own randomness, in the two
cases given that release time
(Proposition~\ref{prop:dnode}, Supplement~\ref{app:proofs}). These inputs can differ only through load.
With the decode instance's admissions taken in random order, the upper bounds of five readings, including a
classifier over all token times, were at most 0.067 under the prediction mode (Supplement~\ref{app:ext}).

\section{Related Work}\label{sec:related}
\noindent\textbf{Timing attacks on prefix caches.}
Direct attacks exploit the timing of the attacker's own requests: Gu et al.\ detected cross-organization cache
sharing in commercial APIs~\cite{gu2025auditing}, and other work reconstructs prompts from the time to first
token~\cite{song2025early,zheng2026inputsnatch,sun2026agent,song2026preserving}. Network observers can also
measure the size and timing of a victim's stream, and timing reveals other inference optimizations as
well~\cite{weiss2024what,mcdonald2025whisper,carlini2024remote}. Indirect attacks instead observe the effects
of shared execution: response order under longest-prefix-match scheduling~\cite{wu2025know,wang2026optileak},
the numerical output of a shared batch~\cite{yona2024stealing,foerster2026quantamination}, or aggregate
KV-cache usage~\cite{wang2026rethinking}. Prior work also detected hits from the timing of other queued requests on
one to four nodes~\cite{prior2026savings}; Section~\ref{sec:measure} tests this on commercial APIs.

\noindent\textbf{Defenses.}
Published defenses restrict which hits occur~\cite{pennas2026cachesolidarity,chu2025selective,wu2026cacheprune,addagada2026governing,pang2024cache,zuo2026privkvsched}
or delay responses~\cite{fonseca2026mitigacao,gu2025auditing,yan2026mirage}, and Section~\ref{sec:priordesign}
analyzes both classes. \sys complements the former: whatever hits a policy permits, it hides their timing
across a cluster.

\noindent\textbf{Cache-aware serving.}
Serving systems reuse prefixes within an engine~\cite{kwon2023efficient,zheng2024sglang}, across engines through
KV tiers and pools~\cite{liu2025lmcache,qin2025mooncake,zuo2025cloudmatrix384}, and across instances through
cache-aware routers~\cite{zhang2026simple,nvidia2026dynamorouter,llmd2026optimized,aibrix2026routing,sglang2026gateway},
some of which identify the timing channel and leave it open~\cite{xia2026skywalker,gim2025pie}. Co-serving
systems fill online steps with offline work up to a latency budget~\cite{chen2026deltaserve,oliaro2026flexllm,yang2026acdc,sun2025hygen,qiao2026conserve},
so their step shapes can still vary with hits, whereas the execution mode reserves exactly the work that each
hit skips. FairInference uses a simulated isolated run to bound a request's delay from above, which still lets
hits produce tokens early~\cite{bali2026token}.

\noindent\textbf{Timing-channel mitigation.}
The prediction mode follows predictive mitigation~\cite{askarov2010predictive} and its per-request form for
interactive systems~\cite{zhang2011predictive}, but derives release times by replaying the scheduler. The
execution mode builds on designs that fix the shape of shared work with dummies or
replicas~\cite{shafiee2015avoiding,ferraiuolo2016lattice,zhou2017camouflage,crooks2018obladi,dauterman2021snoopy,li2014stopwatch,devriese2010noninterference},
but it executes each miss-case step shape once and fills the freed slots with other requests' work. Like
Pacer~\cite{mehta2022pacer}, which conforms outgoing network traffic to a cover schedule, \sys controls only
what the client sees. Because attackers can achieve co-residency in public clouds at a
cost~\cite{ristenpart2009hey,varadarajan2015placement,zhao2024everywhere,shao2026close}, we test load visibility
empirically rather than assume that attacker and victim never share an instance (Section~\ref{sec:measure}). Memory
deduplication offers a precedent: after timing exposed shared pages, hypervisors stopped merging them across
VMs~\cite{vmware2014tps,suse2015cain,razavi2016flip}.

\section{Conclusion}\label{sec:conclusion}
\sys applies least privilege to prefix caching: each request has its own sharing group by default, yet
requests in different groups can still reuse cached computation. The gateway routes each request's miss
twin using only visible hits, and the prediction or execution mode then uses the twin to set when the
response is released.
This design targets deployments with low load visibility: on all tested commercial APIs, our requests could
not tell another request's long prompt from a short one. On our testbed, direct
probes estimated an advantage of at most 0.07 for both modes, against up to 0.98 under unprotected sharing,
and at saturation the prediction mode retained 0.984 of unprotected throughput.

  \makeatletter
  {\edef\@currentlabel{\theproposition}\label{main:last-proposition}}
  \makeatother

  \bibliographystyle{plain}
  \bibliography{bib/refs}
\fi

\ifIncludeSupplement
  \ifIncludeMain
    \clearpage
    \section*{Supplementary Material}
  \else
    \setcounter{proposition}{\getrefnumber{main:last-proposition}}
  \fi
  \appendix
  \renewcommand{\thesection}{S\arabic{section}}
  \renewcommand{\theHsection}{S\arabic{section}}
  \setcounter{table}{0}
  \renewcommand{\thetable}{S\arabic{table}}
  \renewcommand{\theHtable}{S\arabic{table}}
  \setcounter{figure}{0}
  \renewcommand{\thefigure}{S\arabic{figure}}
  \renewcommand{\theHfigure}{S\arabic{figure}}
  \section{Proofs}\label{app:proofs}
We condition all distributions on the information $I$ common to both cases, write $\mathrm{TV}$ for total
variation distance, and let the defense draw randomness independently of the case.

\noindent\textbf{Theorem~\ref{thm:lv}.}
The triangle inequality through $O_0$ gives the first bound, both conditional on $I$ and after averaging
over $I$, as in the test of Section~\ref{sec:measure}.
For the second bound, fix a coupling of $(Z_w,O_w)$ and $O_0$ with the prescribed marginals for each
$w \in \{\mathrm{h},\mathrm{m}\}$. By the triangle inequality, the distance between the joint views
$(Z_{\mathrm{h}},O_{\mathrm{h}})$ and $(Z_{\mathrm{m}},O_{\mathrm{m}})$ is at most
\begin{multline*}
\mathrm{TV}\bigl((Z_{\mathrm{h}}, O_{\mathrm{h}}), (Z_{\mathrm{h}}, O_0)\bigr)
+ \mathrm{TV}\bigl((Z_{\mathrm{h}}, O_0), (Z_{\mathrm{m}}, O_0)\bigr)\\
+ \mathrm{TV}\bigl((Z_{\mathrm{m}}, O_0), (Z_{\mathrm{m}}, O_{\mathrm{m}})\bigr).
\end{multline*}
In each outer term the coupled pairs differ only when $O_w \ne O_0$, so each term is at most
$\Pr[O_w \ne O_0]$, which is at least $\Lambda_w$. Because $O_0$ has the same distribution in both
couplings, the middle term is the mean distance between the release distributions conditional on $O_0$,
which is at most $\eps$ by hypothesis.

The conditional bound holds if $I$ determines $O_0$, or if $O_0$ and $Z_w$ are independent given $I$ in
each coupling, so it must be checked for each mode. Sharing the service's randomness can keep the outer
terms small, but the same randomness also affects the ready time of $x$, and the following example shows
what can then go wrong. Let $O_{\mathrm{h}}=O_{\mathrm{m}}=O_0$ be a fair bit $J$, and set
$Z_{\mathrm{h}}=J$ and $Z_{\mathrm{m}}=1-J$; each release alone is then a fair bit, yet the joint views
are disjoint.

The second bound applies to any classifier that reads the \emph{listeners}, the other requests whose timing the attacker observes, and the probe's release time jointly.
For $K$ listeners under one coupling, let $E_{w,k}$ denote a change in listener~$k$'s observation between
cases~$w$ and~$0$. A union bound then gives $\Pr[O_w \ne O_0] \le \sum_k \Pr[E_{w,k}]$, whereas summing
the individual distinguishing advantages need not bound the joint advantage.
The coupling also covers adaptively chosen routes when the attacker uses the same strategy and randomness in
both cases, because until the first differing observation the attacker sees the same history and acts
identically. If the route also depends on the probe's release time, we supply that time to case~$0$ as
an external input, so that $O_0$ depends on the case only through the release time. By data processing,
the middle term then remains at most $\eps$, and the theorem's first bound gains an additive $\eps$.

\noindent\textbf{Proposition~\ref{prop:reduce}.}
Let $D$ denote the gateway's decisions about $x$. Because they depend only on common inputs and the
defense's own randomness, $D$ has the same distribution $p$ in both cases. Even if the attacker observes
$D$, the distance between the joint distributions of $D$ and the release time is
$\sum_d p(d)\,\mathrm{TV}\bigl(Z_{\mathrm{h}} \mid d,\ Z_{\mathrm{m}} \mid d\bigr) \le \eps$,
since each term is at most $\eps$ by assumption.

\noindent\textbf{Theorem~\ref{thm:target}.}
\emph{Only if.}
If $\phi_w(y) \ge y$ and $\phi_w(Y_w)$ has distribution $Q$, then
$Q(t) = \Pr[\phi_w(Y_w) \le t] \le \Pr[Y_w \le t] = F_w(t)$ for every $t$. Thus $Q$ stochastically
dominates $F_w$, equivalently $Q^{-1}(u) \ge F_w^{-1}(u)$ for all $u$~\cite{shaked2007stochastic}.
\emph{If.}
Let $F_w(y^-)$ be the left limit at $y$ and let $\nu$ be an independent uniform draw on $(0,1)$.
Then $U_w = F_w(Y_w^-) + \nu\,\bigl(F_w(Y_w) - F_w(Y_w^-)\bigr)$ is uniform and
$Y_w = F_w^{-1}(U_w)$ almost surely. We set $Z_w=Q^{-1}(U_w)$, which has distribution $Q$ and, by
\eqref{eq:dominance}, satisfies $Z_w \ge F_w^{-1}(U_w) = Y_w$.
The right-hand side of~\eqref{eq:dominance} is nondecreasing and left-continuous, so it defines the
quantile function of a feasible distribution $Q^{\ast}$, which every feasible $Q$ dominates.
Hence $Q^{\ast}$ minimizes expected cost for any cost nondecreasing in release time.

\noindent\textbf{Inequality~\eqref{eq:main}.}
We couple the cases on the same gate $V$ while preserving each case's joint law of ready time and gate,
which is possible because $V$ depends only on common inputs and the defense's randomness.
The releases then coincide if the ready times coincide or if both are at most $V$, so
$\mathrm{TV}(Z_{\mathrm{h}}, Z_{\mathrm{m}}) \le \Pr[Z_{\mathrm{h}} \ne Z_{\mathrm{m}}]$,
and~\eqref{eq:main} bounds the latter probability. For $k$ probes that do not affect one another's timing,
a union bound over differing releases gives the sum of their $k$ bounds.

\noindent\textbf{Proposition~\ref{prop:gate}.}
We measure release times relative to the twin's predicted ready time~$B$, which is common to both cases,
and let $r_0$ and $c$ be the median and $1-\eps$ quantile of the continuous miss distribution $F$.
The gate offset $\gamma$ has distribution function $G=\eps/(1-F)$ on $[r_0,c]$, with mass $2\eps$
at $r_0$ and $G(c)=1$. Because every hit is ready before $r_0$, its release has distribution $G$.
A miss is released at $\max(\rho,\gamma)$ for $\rho\sim F$ independent of $\gamma$, so its distribution
function is $FG$ on $[r_0,c]$ and $F$ above $c$.
On $[r_0,c]$ the two distribution functions therefore differ by the constant $G-FG=(1-F)G=\eps$, and they
assign the same mass to every interval inside $(r_0,c]$.
At $r_0$ the hit distribution has mass $2\eps$, whereas the miss distribution has mass $F(r_0)\,2\eps=\eps$.
Above $c$, only the miss distribution has mass, exactly $1-F(c)=\eps$, so the total variation distance
is $\eps$.

Now suppose that the gate uses an estimate of $F$ and that hits are still ready before its lower endpoint.
For a true miss distribution $F$ with density $f$, the miss release density exceeds the hit release
density inside the gate's range only where the true hazard rate $f/(1-F)$ exceeds the estimated rate.
If the true rate stays at or below the estimate throughout that range, the only excess miss mass
therefore lies above $B+c$. The randomized gate and a fixed gate at $B+c$ then leak that same tail mass.

\noindent\textbf{Proposition~\ref{prop:exec}.}
By the triangle inequality,
$\mathrm{TV}(Z_{\mathrm{h}}, Z_{\mathrm{m}}) \le \mathrm{TV}(Z_{\mathrm{h}}, \tilde{Y}) + \mathrm{TV}(\tilde{Y}, Y_{\mathrm{m}})$.
In the hit case, $Z_{\mathrm{h}} = \max(Y_{\mathrm{h}}, \tilde{Y})$ differs from $\tilde{Y}$ only when
$Y_{\mathrm{h}} > \tilde{Y}$, so the first term is at most $\Pr[Y_{\mathrm{h}} > \tilde{Y}]$.
If $\tilde{Y}=Y_{\mathrm{m}}+\Delta$ for constant $\Delta$ and $Y_{\mathrm{m}}$ has a unimodal density
with maximum $f^{\ast}$, the second term equals $\Pr[a < Y_{\mathrm{m}} \le a + |\Delta|]$ for some $a$
and is therefore at most $|\Delta| f^{\ast}$. For a normal distribution with standard deviation
$\varsigma$, $f^{\ast} \approx 0.4/\varsigma$.

\noindent\textbf{Disaggregated decode.}
For request~$x$, let $T_1$ denote the first-token release time and $\mathbf{T}$ the later token times,
and let $\Xi$ collect other arrivals at the decode instance and the instance's own randomness.

\begin{proposition}\label{prop:dnode}
If, given $T_1$ and $\Xi$, the distribution of $\mathbf{T}$ is the same in both cases, then the advantage on the
view $(T_1, \mathbf{T})$ is at most
$\mathrm{TV}(T_{1,\mathrm{h}}, T_{1,\mathrm{m}}) + \mathbb{E}_{t \sim T_{1,\mathrm{h}}}\,
\mathrm{TV}\bigl(\Xi_{\mathrm{h}} \mid t,\ \Xi_{\mathrm{m}} \mid t\bigr)$.
\end{proposition}

\noindent
Both cases generate $\mathbf{T}$ from $(T_1,\Xi)$ through the same kernel, so by data processing the
distance on $(T_1,\mathbf{T})$ is at most the distance on $(T_1,\Xi)$. The chain rule for total
variation bounds the latter by the stated sum.

In \sys, the premise holds up to the numerical differences that Section~\ref{sec:goals} excludes, and
$\Xi$ can differ between the cases only through load, because a hit frees prefill work and can shift
when other requests' first tokens are released.

\section{Consistency Check, Bait Requests, and Full Readings}\label{app:cdt}
\noindent\textbf{Consistency check.}
The harness runs the scheduler class that vLLM loads and the gateway's reservation logic on one instance.
A model engine advances in steps whose duration depends only on their shape, and each request ends at its
first token. Every scenario fixes the arrivals, prompts, and minimum dispatch delay and runs twice, once
with $x$'s prefix cached outside its group and once without. We compare, for every request, the step and
time of its first token, its output length and finish reason, and the cached tokens that it reports. One
set of scenarios enumerates a regular grid of arrivals with a dispatch delay of 250\,ms, and the others
draw arrivals at random with delays of 0, 250, or 500\,ms. In some of the random sets, the probe arrives
at a random time within 6\,s of the victim's dispatch, including during the victim's prefill.

With slots filled only by dummy tokens, the two runs agreed for every request in 5{,}673 enumerated and
7{,}000 random scenarios, and no request reported cached tokens. With substitutes and reservation as
deployed, no request reported cached tokens either, and $x$'s first token differed in 28 of 9{,}673
scenarios: 21 of 5{,}673 enumerated ones, 2 of 3{,}000 random ones, and 5 of 1{,}000 with random probe
arrival. In all 28, $x$ was later in the hit case, by one to three steps. The 21 enumerated scenarios share
one cause. The gateway reserved a request whose own dispatch time equaled the dispatch time of $x$, so the
instance did not set it aside, and it entered the queue ahead of $x$ in the hit case and behind it in the
miss case. In our two-instance security runs and open-loop runs, the numbers of reserved requests and of
requests set aside were equal in every run, so no reservation had a zero wait there. In three scenarios, $x$
arrived while the victim's prefill was running. The hit covered only part of the prefix when the slot was
built, a substitute was chosen for that part, and one step later the hit had grown, so the slot was
rebuilt with dummy tokens and the substitute continued as an ordinary request ahead of $x$. The same
scenarios agreed when slots held only dummy tokens. Our probes arrive after the victim's first token and
do not exercise this situation. In the last four scenarios, the observed request was itself drawn as a substitute
for another slot, directly or within a chain of substitutes, and its first token came one step later in
the hit case. The
scheduler class for mixture-of-experts models (Section~\ref{sec:content}) matched the deployed class
field by field on these scenarios. The check covers the scheduling logic of one instance. It does not
cover measured step times, routing across instances, or tokens after the first.

\begin{table}[t]
\centering\small
\setlength{\tabcolsep}{3.5pt}
\caption{Bait requests on two instances: single-probe advantage from the bait's first-token time, with its
upper bound and noise bound. Delay is the time from the probe to the bait, Reserved is the share of
hit-case baits that the gateway reserved for their own probe's slot, and no miss-case bait was reserved
for its probe. Light load pools three seeds and full load six.}
\label{tab:bait}
\begin{tabular*}{\columnwidth}{@{\extracolsep{\fill}}llrrrrr@{}}
\toprule
Load, delay & Hit & Probes & Reserved & Adv. & Upper & Noise \\
\midrule
Light, 0.05\,s & 256  & 1{,}268 & 0     & $-0.005$ & 0.046 & 0.077 \\
               & 2048 & 1{,}301 & 0     & $-0.017$ & 0.038 & 0.076 \\
               & 7168 & 1{,}246 & 0     & 0.012    & 0.062 & 0.078 \\
Full, 0.30\,s  & 256  & 2{,}627 & 0.033 & 0.012    & 0.046 & 0.053 \\
               & 2048 & 2{,}589 & 0.031 & 0.027    & 0.060 & 0.054 \\
               & 7168 & 2{,}578 & 0.010 & 0.010    & 0.046 & 0.054 \\
Full, 0.24\,s  & 256  & 2{,}620 & 0.067 & 0.012    & 0.042 & 0.053 \\
               & 2048 & 2{,}648 & 0.059 & 0.012    & 0.048 & 0.053 \\
               & 7168 & 2{,}677 & 0.014 & 0.011    & 0.044 & 0.053 \\
\bottomrule
\end{tabular*}
\end{table}

\noindent\textbf{Bait requests.}
Each trial adds a bait to the probe trial of Section~\ref{sec:evalsetup}: the attacker sends a fresh
prompt, longer than the guessed prefix, 0.05 to 0.3\,s after the probe, and we classify the bait's
first-token time as we do the probe's. Table~\ref{tab:bait} gives every cell, with the smaller number of
probes of the two cases. At light load the gateway applies no dispatch delay, holds no request, and
reserved no bait. At full load, where the dispatch delay is 0, 500, or 250\,ms by load block, it reserved
27 to 182 of about 2{,}700 hit-case baits per cell for their own probe's slot. One upper bound exceeded
its noise bound, for 2{,}048-token hits at a delay of 0.3\,s. Without the reserved baits, that cell read
0.034 with an upper bound of 0.064, and the other five full-load cells read 0.006 to 0.023, so the
reserved baits did not produce these readings. Within the load block where reservation was possible at
this delay, the 32 to 54 reserved baits per three-seed half for 256- and 2{,}048-token hits differed from
the miss-case baits by $-232$ to $+61$\,ms at the median, with no consistent sign. Read as two halves of
three seeds, three of the twelve full-load cells exceeded their noise bounds, and none did so in both
halves.

\begin{table}[t]
\centering\small
\setlength{\tabcolsep}{3.5pt}
\caption{All cells of Table~\ref{tab:sec} and those for 2{,}048-token hits: probes per case, single-probe
advantage, its one-sided 95\% upper bound, and the noise bound at that number of probes.}
\label{tab:secfull}
\begin{tabular*}{\columnwidth}{@{\extracolsep{\fill}}lrrrrrr@{}}
\toprule
Arm & $M$ & Hit & Probes & Adv. & Upper & Noise \\
\midrule
Unprotected & 2 & 2048 & 1{,}524 & $-0.006$ & 0.045 & 0.070 \\
            &   & 7168 & 1{,}517 & 0.210    & 0.239 & 0.070 \\
            &   & 16384 & 132    & 0.980    & 0.994 & 0.237 \\
            & 4 & 2048 & 2{,}758 & $-0.003$ & 0.028 & 0.052 \\
            &   & 7168 & 2{,}764 & 0.184    & 0.209 & 0.052 \\
            &   & 16384 & 130    & 0.937    & 0.969 & 0.239 \\
            & 8 & 2048 & 7{,}660 & 0.002    & 0.022 & 0.032 \\
            &   & 7168 & 7{,}786 & 0.170    & 0.182 & 0.032 \\
            &   & 16384 & 114    & 0.929    & 0.963 & 0.256 \\
\addlinespace[2pt]
Prediction  & 2 & 2048 & 8{,}033 & $-0.009$ & 0.015 & 0.031 \\
            &   & 7168 & 8{,}148 & 0.000    & 0.021 & 0.031 \\
            &   & 16384 & 242    & $-0.016$ & 0.096 & 0.175 \\
            & 4 & 2048 & 2{,}592 & $-0.006$ & 0.030 & 0.053 \\
            &   & 7168 & 2{,}529 & 0.002    & 0.045 & 0.054 \\
            &   & 16384 & 243    & $-0.004$ & 0.104 & 0.175 \\
            & 8 & 2048 & 7{,}436 & $-0.001$ & 0.019 & 0.032 \\
            &   & 7168 & 7{,}216 & $-0.006$ & 0.019 & 0.033 \\
            &   & 16384 & 229    & 0.068    & 0.175 & 0.180 \\
\addlinespace[2pt]
Execution   & 2 & 2048 & 8{,}816 & $-0.004$ & 0.017 & 0.029 \\
            &   & 7168 & 8{,}952 & $-0.001$ & 0.020 & 0.029 \\
            &   & 16384 & 143    & 0.006    & 0.162 & 0.228 \\
            & 4 & 2048 & 2{,}075 & $-0.004$ & 0.045 & 0.061 \\
            &   & 7168 & 2{,}138 & 0.003    & 0.043 & 0.060 \\
            &   & 16384 & 168    & $-0.030$ & 0.131 & 0.211 \\
            & 8 & 2048 & 6{,}088 & 0.003    & 0.028 & 0.036 \\
            &   & 7168 & 6{,}202 & $-0.006$ & 0.021 & 0.035 \\
            &   & 16384 & 160    & $-0.020$ & 0.133 & 0.216 \\
\bottomrule
\end{tabular*}
\end{table}

\noindent\textbf{Full readings.}
Table~\ref{tab:secfull} lists every cell of Table~\ref{tab:sec} and the cells for 2{,}048-token hits, with
the smaller number of probes of the two cases. The estimate can be negative because the threshold is scored
on held-out probes. Read separately for the three load blocks, for 2{,}048- and 7{,}168-token hits, the
prediction mode's 18 estimates ranged from $-0.021$ to 0.025 and
the execution mode's from $-0.014$ to 0.033. All of the prediction mode's block upper bounds and 17 of the
execution mode's were at or below their noise bounds. The exception was 7{,}168-token hits on two
instances in the block without a dispatch delay, with an estimate of 0.033 and an upper bound of 0.057
against 0.052. Under unprotected sharing, the block estimates for 2{,}048-token hits were 0.011 to 0.049,
although the pooled estimates were near zero, because first-token times differ between load blocks by far
more than a short hit saves.

An attacker could also try to remove the effect of load by adjusting each probe's time by the median of
the probes in the same 30\,s window that did not overlap it. On 7{,}168-token hits, this gives upper
bounds of 0.022 for the prediction mode and 0.021 for the execution mode, against 0.446 under
unprotected sharing.

\section{Total Variation and the Threshold Classifier}\label{app:tvthr}
Definition~\ref{def:secure} bounds a total variation distance, whereas Section~\ref{sec:evalsetup} reports
the advantage of one threshold on first-token time. This section explains both choices with the probes of
the security runs: twelve cells of the two modes, six cells of unprotected sharing, and six cells of
the no-reuse baseline, which has no hits.
Because the threshold estimator uses only the ranks of the times, the distribution of its upper bound
without any effect depends only on the number of probes per case, so one simulated noise bound per sample
size applies to every cell.

\begin{table}[t]
\centering\small
\setlength{\tabcolsep}{2.5pt}
\caption{Plug-in estimates of total variation distance with $B$ equal-mass bins, for 7{,}168-token hits
in the execution mode on two instances. Halves splits the miss-case probes into random halves of $n=4{,}508$ and gives
the mean estimate over 200 splits, followed by its approximation. Hit, miss compares the 8{,}952 hit-case
probes with the 9{,}017 miss-case probes, and Permuted is the 95th percentile of that estimate under
permuted labels.}
\label{tab:tvbias}
\begin{tabular*}{\columnwidth}{@{\extracolsep{\fill}}lrrrrrrr@{}}
\toprule
$B$ & 2 & 8 & 32 & 128 & 512 & 2048 & none \\
\midrule
Halves              & 0.008 & 0.023 & 0.047 & 0.094 & 0.189 & 0.375 & 1.00 \\
$\sqrt{B/(\pi n)}$  & 0.012 & 0.024 & 0.048 & 0.095 & 0.190 & 0.380 & --- \\
Hit, miss           & 0.007 & 0.019 & 0.038 & 0.063 & 0.137 & 0.276 & 1.00 \\
Permuted            & 0.015 & 0.024 & 0.041 & 0.075 & 0.142 & 0.281 & --- \\
\bottomrule
\end{tabular*}
\end{table}

\noindent\textbf{Estimating the distance directly.}
The plug-in estimate sorts the times of both cases into $B$ bins of equal mass and takes half the summed
absolute difference between the two cases' bin frequencies. This estimate is biased upward. For two
samples of $n$ probes from one distribution it is about $\sqrt{B/(\pi n)}$ rather than zero, and without
bins it equals 1, because no two times coincide (Table~\ref{tab:tvbias}). An estimate therefore has
meaning only relative to its value under permuted labels, and it is the advantage of a histogram
classifier scored on the probes that trained it. Compared with the 95th percentile under
permuted labels at the six bin counts of Table~\ref{tab:tvbias}, the estimate was higher in 4 of 72
comparisons for the two modes and in 0 of 36 under no reuse, against 25 of 36 under unprotected sharing.
We therefore report the advantages of classifiers on held-out probes, each of which is a lower bound on
the distance.

\noindent\textbf{Richer classifiers.}
A threshold detects a hit that moves first-token times in one direction, but a hit could also change
their shape. We therefore compared seven classifiers: the threshold, an interval between two thresholds,
histograms with 4, 8, 16, and 32 equal-mass bins that guess a hit in the bins where hits were more
frequent, and a ratio of kernel density estimates. Each is fitted on half of a cell's probes and scored
on the other half, as in Section~\ref{sec:evalsetup}, and 1{,}000 label permutations give a one-sided
$p$-value. Of the 84 tests on the two modes' cells, 3 had $p<0.05$, compared with 2 of 42 under no reuse;
about 4 and 2 are expected without any effect. The three were histograms with 32 or 8 bins, with
advantages of 0.007, 0.020, and 0.033.

Under unprotected sharing, all seven classifiers detected 7{,}168-token hits at every cluster size, and
the threshold reached 0.97 to 1.00 of the best classifier's advantage. These analyses draw their own
random halves, so the threshold's values differ slightly from Table~\ref{tab:secfull}. The 2{,}048-token hits are the
exception. There the threshold read $-0.010$, $-0.008$, and 0.003 on two, four, and eight instances,
whereas the 32-bin histogram read 0.008, 0.024, and 0.011, with $p$-values of 0.13, 0.012, and 0.017, and
no other classifier had $p<0.05$. Without a defense, a short hit thus changes the shape of the
distribution more than its location, and the threshold understates what it reveals.

\noindent\textbf{Sensitivity.}
To see which differences these classifiers resolve, we split the miss-case probes of
Table~\ref{tab:tvbias} into random halves, altered one half, and counted how often each classifier's
advantage exceeded the 95th percentile of unaltered splits, in 200 trials. Widening the spread around the
median by a factor of 1.10 was detected in 0.94 of trials by the threshold and in every trial by the
other six classifiers. A shift of 20\,ms was detected in only 0.07 to 0.14 of trials by any classifier,
and moving a random tenth of the probes 150\,ms earlier and another tenth 150\,ms later in 0.02 to 0.07,
close to the 0.05 expected without a change. At this sample size, none of the seven classifiers resolves
differences of the last two kinds.

\section{Release Rules Compared}\label{app:rules}
The randomized gate of Section~\ref{sec:gate} trades some delay for robustness. Its target is slightly
later than $Q^{\ast}$, so misses can wait, but both cases use the same release rule, which needs no
estimate of the hit distribution. Because all ready times below the gate map to a single release time,
errors in $F$ mainly affect delay. With every request isolated, a first-order estimate from the unprotected
runs of Section~\ref{sec:eval} puts the mean delay of the ideal rule of Section~\ref{sec:instance} for hits
at 255 to 321\,ms, and on two workloads without probes, the prediction mode held hits for 278 to 340\,ms on
average.

A \emph{quantile band} modifies the transport map of Section~\ref{sec:instance} by randomizing each
request's quantile within a band of width~$w$. Compared with the randomized gate of
Section~\ref{sec:gate}, it merges fewer ready times and holds misses for less time.
Table~\ref{tab:rules} compares the three rules in replay on the security runs of Section~\ref{sec:evalsec}
and gives the gate's online results on the same runs; all three rules held hits for 0.38\,s on average.

Across 65{,}876 steps on two L40S instances, the step-duration model of Equation~\eqref{eq:steplaw} has a
residual standard deviation of 3.34\,ms, which the online correction factor per instance reduces to 2.2 to
2.4\,ms. For a miss that runs on $S$, the request is its own twin, and the prediction identifies its first
step in 98.5 to 98.9\% of misses and is one step early in the rest.
To estimate these distributions, the deployed gate groups miss residuals by predicted time~$B$ and prompt
length, and hit residuals by the number of tokens left to compute. It keeps the latest 400 samples per
class and uses a pooled distribution for classes with fewer than 50 samples.

\noindent\textbf{The cross-instance check.}
In six unprotected runs on two workloads, 9 to 38\% of cross-instance hits became ready after their
twin's predicted time on~$S$, and reducing this fraction to 1\% would require a margin of 0.37 to 1.01\,s.
The gateway's check instead uses a cross-instance hit only if its predicted ready time on~$R$ plus the
0.75 quantile of the hit residual is no later than the twin's predicted time on~$S$ plus the 0.25 quantile
of the miss residual.

\noindent\textbf{A floor on the gate's width.}
The prediction mode cannot observe the timing jitter of the twin's unexecuted steps, so its gate must be
wide enough to absorb that jitter. Under load, with errors in the predicted shape, the standard deviation
of a full step reaches 10\,ms, and the 0.99 quantile of four steps together lies up to 42\,ms above their
median. The execution mode removes this floor on gate width by running the steps.

\section{Delay Rules of Prior Defenses}\label{app:base}
We replay delay rules with a common target (Section~\ref{sec:priordesign}) on the unprotected security runs, and
Table~\ref{tab:base} reports the smallest parameter on our grid that brings every cell's upper bound
to its noise bound or below, 0.071 for 2{,}048-token hits and 0.070 for 7{,}168-token hits. A fixed target releases each first token at arrival plus $T$ or when ready, whichever
is later, and quantized release rounds the ready time up to the next multiple of a quantum.
Per-load-block padding instead releases at the $q$ quantile of latency in the request's load block, one of three groups of load stages (Supplement~\ref{app:evaldetail}).

Three further rules delay only hits. Padding each hit to the miss $q$ quantile within its load block
reaches a best upper bound of 0.155, conditioning further on length gives 0.152, and adding exponential
noise gives 0.124, all above the noise bounds. At some settings, a padded hit even arrives later than its miss twin, which raises
the bound in the otherwise weak 2{,}048-token cell to 0.41--0.50. The parameters in Table~\ref{tab:base}
hold over a whole run, whereas keeping every load block of eight runs within its noise bound would require
$T=5$\,s and $q=1.0$.

The crossing rate is the fraction of hits ready after their target release time, and at comparable rates
of 5 to 9\%, finer timing information reduces mean delay. A single delay for all requests adds 1.95\,s,
a delay per load block adds 0.83\,s, and the twin's prediction adds 0.17\,s, while requests without hits
wait 0.014\,s under the twin's prediction.

\begin{table}[t]
\centering\small
\setlength{\tabcolsep}{3pt}
\caption{Offline replay of prior release rules. For each rule we give the smallest tested parameter that
brings the upper bound to its noise bound or below in every cell of the unprotected security runs. Added delay for
non-probe requests is the mean and 90th percentile in security runs and the mean in runs without probes.
Hit-only rules reach no bound below 0.124, and the last row gives the prediction mode's measured mean
hold over all requests.}
\label{tab:base}
\begin{tabular}{@{}lccc@{}}
\toprule
Rule & Parameter & Added delay & No probes \\
\midrule
Fixed target & $T = 4.5$\,s & 2.26 / 3.44\,s & 3.12\,s \\
Quantized release & quantum 5\,s & 2.78 / 3.95\,s & 3.62\,s \\
Pad all, per load block & $q = 0.975$ & 0.98 / 1.65\,s & 0.60\,s \\
Only hits, 3 rules & none & --- & --- \\
Prediction, hold & --- & --- & 0.20\,s \\
\bottomrule
\end{tabular}
\end{table}

\section{Prior Defenses and the Routing Ablation}\label{app:prior}
\begin{table}[t]
\centering\small
\caption{Published defenses and \sys on two instances: upper bounds of the single-probe advantage for
7{,}168-token hits and mean delay added per request. \emph{Later} probes follow the first cross-account
reuse of the prefix under CacheSolidarity and the account's first request on it under Mirage. Padding rules are replayed on unprotected
runs, the rows for \sys repeat Table~\ref{tab:secfull}, and the execution mode's cost appears as
throughput (Section~\ref{sec:evalperf}).}
\label{tab:prior}
\setlength{\tabcolsep}{3pt}
\begin{tabular*}{\columnwidth}{@{\extracolsep{\fill}}lrrr@{}}
\toprule
 & \multicolumn{2}{c}{Upper bound} & Delay \\
\cmidrule(lr){2-3}
Defense & First probe & Later & in s \\
\midrule
Unprotected          & \multicolumn{2}{c}{0.302} & -- \\
CacheSolidarity      & 0.262 & 0.052 & -- \\
Mirage               & 0.257 & 0.029 & 1.60 \\
Padding to a target  & \multicolumn{2}{c}{$\le 0.058$} & 0.98--2.78 \\
\addlinespace[2pt]
Prediction           & \multicolumn{2}{c}{0.021} & 0.17 \\
Execution            & \multicolumn{2}{c}{0.020} & -- \\
\bottomrule
\end{tabular*}
\end{table}
\noindent\textbf{Setup.}
Except for the joining-instance runs, the runs in this section used nine closed-loop load
stages with 6 to 19 requests in flight per instance, and on one instance the router served a single
engine. The noise bounds for Table~\ref{tab:prior} are 0.106 for unprotected sharing, 0.073 and 0.068 for
CacheSolidarity's first and later probes, 0.042 for Mirage, 0.070 to 0.071 for padding to a target, 0.031
for the prediction mode, and 0.029 for the execution mode.

\noindent\textbf{CacheSolidarity.}
CacheSolidarity marks the point where a prefix is first reused across accounts and lets other
accounts reuse it only up to that point. When the probe is itself the first cross-account reuse of the
victim's prefix, it is served in full: on two instances, such probes gave 0.229 for 7{,}168-token hits,
with an upper bound of 0.262 for 1{,}408 probes, beside 0.253 and 0.302 for the unprotected run measured
alongside. For the later probes of Table~\ref{tab:prior}, a 512-token template shared by benign accounts
was marked before the victim's request, so a probe from a new account may reuse the template but
recomputes the victim's 7{,}168 tokens. The router still matches the probe against all cached blocks: on
two instances, it sent 98\% of hit-case probes to the victim's instance, against 73\% of miss-case
probes. On one instance, the upper bound for these probes was 0.046.

\noindent\textbf{Routing ablation.}
The ablation of Section~\ref{sec:evalabl} keeps the prediction mode's rule for moving a request to $R$
(Section~\ref{sec:sharegate}) but times release by a twin on the instance where the request runs. On one
instance, it schedules every request exactly as the prediction mode does, so one run serves both, with
an upper bound of 0.043 for 7{,}168-token hits. Its probes, like all probes in the joining-instance
runs, came from the victim's account, and probes from the victim's account use no template. On two
instances under steady load, it gave an upper bound of 0.075 against a noise bound of 0.082, beside 0.069
for the prediction mode.

\noindent\textbf{Controls and short hits.}
The unprotected runs measured alongside the ablation and CacheSolidarity sent probes from new
accounts, which reuse the victim's prefix without a defense just as the victim's own account would. They
gave upper bounds of 0.316 and 0.334 for 7{,}168-token hits on one instance and 0.268 and 0.302 on two,
and Table~\ref{tab:prior} shows the two-instance run alongside CacheSolidarity. For 2{,}048-token hits, their point
estimates were at most 0.068, so short hits test the defenses only weakly: the ablation gave upper
bounds of 0.062 on one instance and 0.052 on two, and CacheSolidarity gave 0.053 and 0.045.
Cache-aware routing concentrates benign requests on one instance, so on two instances the template was
sometimes evicted from the other. Restricting CacheSolidarity's single probes to trials in which the
template had been marked before the probe changed the upper bound from 0.052 to 0.058.

\noindent\textbf{Routing under steady load.}
A miss usually ran on the victim's instance too, because the victim's long request had diverted
concurrent requests to the other instance. Only 29\% of hit-case probes under the ablation went to
an instance other than the one a miss would use, and these probes alone gave a point estimate of 0.056.
Under CacheSolidarity, probes whose miss would run on the other instance gave 0.091, with an upper bound
of 0.157 for 456 to 494 probes per case, within the noise bound of 0.172 for that count, and the
remaining probes gave 0.013, with an upper bound of 0.064.

\noindent\textbf{Joining instance.}
Each 5\,s cycle took the second instance out of routing for its first 3\,s. The victim's request ran on
the first instance in every trial, none of the roughly 15{,}000 requests per run that arrived while the
second instance was out was routed to it, and every probe was sent as the second instance returned. The
second instance kept its cache between cycles; for the probe, what mattered was that it held none of the
victim's prefix and carried almost no load when it returned. Four
1{,}500\,s stages held 2, 3, 4, and 5 requests in flight in the cluster, a lighter load than in the other
runs, with 585 and 623 probes per case. The ablation gave point estimates of 0.39 to 0.46 in every
stage and an upper bound of 0.433 overall, and the prediction mode gave about $-0.01$ in each stage and an
upper bound of 0.074, against a noise bound of 0.116. In two further runs of this setting under
the ablation, each trial also drew whether to probe 0.2\,s before the second instance returned,
when only the first instance could take the probe, and all such probes ran on it. Probes sent as the
second instance returned gave 0.393, with an upper bound of 0.437 against a noise bound of 0.116 for 568
and 630 probes per case, and probes sent 0.2\,s earlier gave 0.061, with an upper bound of 0.112
against a noise bound of 0.103 for 609 probes per case. The difference is 0.33, with a 95\% interval
of 0.26 to 0.42. We did not run unprotected sharing or CacheSolidarity in this setting.

\noindent\textbf{Mirage.}
We replayed Mirage's padding on seven unprotected two-instance runs, with the parameters from its paper
and our own values for those it leaves open, fitting the miss latency per instance over the 256 most
recent requests without cached tokens. Padded probes gave upper bounds of 0.029 and 0.020 for 7{,}168-
and 2{,}048-token hits, within the noise bound of 0.042, and windows of 64 and 1{,}024 requests gave at
most 0.029. Each account's unpadded first probe gave 0.241 for 7{,}168-token hits, with an upper bound of
0.257, far above that noise bound. Padding every prefix would add 1.60\,s to the first token on average
and 2.47\,s at the 90th percentile, or 1.52\,s and 1.44\,s on average with windows of 64 and 1{,}024,
whereas the prediction-mode runs measured alongside four of these runs held non-probe requests for 0.164
to 0.167\,s on average.

\section{Mixture of Experts, Multi-GPU Instances, and Disaggregated Decode}\label{app:ext}
\noindent\textbf{Mixture of experts.}
We tested Qwen3-30B-A3B with FP8 weights on two L40S instances at 2 and 8 requests in flight and at full
load, with 7{,}168-token probe prefixes of natural text, code, or a repeated token. In the prediction
mode's runs, each account formed a sharing group and each probe used a fresh account, so all of a probe's
hits were hidden. With the change of Section~\ref{sec:content}, its single-probe upper bounds stayed at or
below their noise bounds in all nine cells, with estimates of $-0.013$ to 0.023, and none exceeded 0.094.

The execution mode's runs isolated every request and drew their other requests from the Mooncake trace.
At 2 requests in flight, unprotected sharing gave advantages of 0.86, 0.83, and 0.87 for natural text, the
repeated token, and code, on 394 to 449 probes per case; 0.86 of its hit-case probes ran on the holder
and read cached tokens. Without the change of Section~\ref{sec:content}, the execution mode gave 0.71,
0.83, and 0.87. With the change, in which the dummy tokens of a slot carry the hit's own token ids, the
estimates at 2 requests in flight were 0.010, 0.009, and 0.034, with upper bounds of 0.068, 0.060, and
0.071 against noise bounds of 0.077, 0.079, and 0.077 on 1{,}200 to 1{,}296 probes per case. At full load
they were $-0.011$, 0.021, and $-0.012$, with upper bounds of 0.051, 0.065, and 0.045 against 0.076, 0.076,
and 0.078. At 8 requests in flight, they were $-0.012$, $-0.012$, and $-0.011$, with upper bounds of
0.026, 0.029, and 0.033 against 0.057, 0.058, and 0.057 on 2{,}241 to 2{,}381 probes per case.
In open-loop runs on the same trace, the execution mode delivered 0.66 to 0.84 of unprotected sharing's
goodput at first-token SLOs of 0.5 to 5\,s, against 0.62 to 0.71 for the no-reuse baseline and 0.88 to
0.99 for the prediction mode, and its highest sustained rate was 0.835 of the unprotected one.

\noindent\textbf{Multi-GPU instances.}
\begin{table}[t]
\centering\small
\setlength{\tabcolsep}{4pt}
\caption{Two Qwen3-32B instances with two tensor-parallel A100 GPUs each, and two Llama-3.1-8B instances
with one L40S each. Spread is the median interquartile range for repeated steps of one shape, relative
to step time. Single-probe upper bounds are for 2{,}048- and 7{,}168-token hits, respectively; the L40S
bounds pool four runs, and the A100 bounds use one run each. Throughput includes probes and is relative
to unprotected sharing, which for the execution mode on A100 was measured in a separate run.}
\label{tab:tp}
\begin{tabular}{@{}lcc@{}}
\toprule
 & TP2, A100 & One L40S \\
\midrule
Step of 4{,}096 tokens & 614.7\,ms & 334.4\,ms \\
Median jump every 128 tokens & 9.64\,ms & 6.01\,ms \\
Spread of one shape & 0.65--0.82\% & 1.16--1.44\% \\
Bounds, prediction mode & 0.038, 0.043 & 0.015, 0.021 \\
Bounds, execution mode & 0.044, 0.043 & 0.017, 0.020 \\
Throughput, prediction mode & 0.989 & 0.967 \\
Throughput, execution mode & 0.691 & 0.737 \\
\bottomrule
\end{tabular}
\end{table}
With tensor parallelism, every layer ends with an all-reduce whose duration may depend on more than step
shape. To test whether step shape still sets step time, we ran two Qwen3-32B instances, each with two A100
GPUs linked by NVLink, on the self-reuse workload with probes (Table~\ref{tab:tp}). We recalibrated the
prediction mode's step-time model but kept the execution mode's L40S timing parameters. Step time retained
its smooth dependence on depth and jumped at all 29 boundaries of the 128-token staircase, and repeated
steps of one shape had a smaller relative spread than on one L40S.

For both modes, all four single-probe upper bounds were at or below the noise bound of 0.069 for 1{,}600
trials per case, and the execution mode's estimates were $-0.006$ and 0.002, whereas unprotected sharing
read 0.276 for 7{,}168-token hits and 0.044 for 2{,}048-token hits. With each probe centered on its 30\,s
block, the execution mode's upper bound for 7{,}168-token hits was 0.057 to 0.072, depending on the
centering statistic, and so straddled the bound.
The shorter hits saved only 0.1 to 0.3\,s of a slower prefill, so they provided a weak test here.
In the execution mode, splitting a step's tokens into two chunks changed its duration by 0.5 to 1\,ms;
we saw the same discrepancy on one A100, which points to the GPU as its source. These single runs on one
node leave pipeline parallelism and cross-node tensor parallelism untested.

\noindent\textbf{Disaggregated decode.}
We used one prefill and one decode instance as in Section~\ref{sec:dnode}, with one sharing group per
account and a fresh account per probe. To isolate decode timing, we randomized admission order across
cases. Across five readings, including a classifier over all token times, the upper bounds were at most
0.067 under the prediction mode.

\section{Execution Mode Under Load}\label{app:execload}
We report \emph{recovery}, the fraction of the throughput gap between the no-reuse baseline and
unprotected sharing that a mode closes. If $T_X$, $T_{\mathrm{n}}$, and $T_{\mathrm{u}}$ denote throughput
under scheme~$X$, the no-reuse baseline, and unprotected sharing, recovery is
$(T_X-T_{\mathrm{n}})/(T_{\mathrm{u}}-T_{\mathrm{n}})$.
The \emph{share of GPU-time savings} measures a different quantity. When steps are full, scheme~$X$ saves
a fraction $(E_{\mathrm{n}} - E_X)/(E_{\mathrm{n}} - E_{\mathrm{u}})$, where $E_{\mathrm{n}}$, $E_X$, and
$E_{\mathrm{u}}$ are the GPU times to serve the same requests under no reuse, $X$, and unprotected sharing.

To see how recovery depends on cluster size, we ran the self-reuse workload without probes on 2, 4, 8, 16,
and 32 instances, of which one was real, two on 32 instances, and the rest emulated, at three middle-load
stages with a minimum dispatch delay of 250\,ms. The execution mode reached 0.486, 0.553, 0.580, 0.608,
and 0.626 of unprotected throughput, against 0.383, 0.456, 0.492, 0.525, and 0.548 for the no-reuse
baseline, so its recovery was 0.167, 0.179, 0.173, 0.174, and 0.173, with one run each. Within each run,
recovery rose with load, from 0.105--0.121 at the lightest of the three stages to 0.221--0.234 at the
heaviest. Letting the gateway choose among 16 rather than 64 queued requests gave a recovery of 0.175 on
16 instances and 0.161 on 32, and a minimum dispatch delay of 500\,ms lowered it to 0.145 and 0.134 on 8
and 16 instances.

\noindent\textbf{Sharing groups.}
In one run on two real instances without probes, we made each conversation a sharing group, so hits
within a conversation were visible and all other hits hidden. In the self-reuse workload, 0.89 of reusable
tokens fall within a conversation. The execution mode kept unprotected sharing's cached share of prompt
tokens, 0.62, and reached 1.015 times its throughput. Dummy tokens made up 0.040 of the tokens it
computed, and it executed one twin in full. Requests with a within-group hit waited 0.1\,ms on average
for gateway dispatch.

One security run on two real instances tested the prediction mode with sharing groups. Group names were
scoped to accounts, so the probe and the victim belonged to different groups even when their group names
matched. The probe either had an empty cache or already held the first 1{,}024 prefix tokens in its
group. In the latter situation, it had a visible hit in both cases and a hidden hit only in the hit case. As
a positive control for sharing within a group, a request in the victim's group reported its hit in the
usage fields whenever it reached the holder. Unprotected sharing gave an advantage estimate of 0.245,
while the prediction mode's single-probe upper bounds were 0.059 to 0.071 on 595 to 774 probes per case.

\section{Full External Measurements}\label{app:measure}
\begin{table*}[t]
\centering\footnotesize
\setlength{\tabcolsep}{3.5pt}
\caption{External measurements. Repeated rows use more burst trials, prefix rows use a cached prefix
shared by A and its listeners, and the last three rows use spaced listeners. A's shift is the difference
in median first-token time, added prompt minus base, with a bootstrap 95\% interval, and A's $\hat\eps$
gives the single-probe advantage estimate and upper bound. \emph{Own} denotes A's key, workspace or
project, with a different cache key per request in the repeated OpenAI row; \emph{other ws} and
\emph{other proj.}\ denote another workspace or project in the same organization or account, and
\emph{other org} a different organization. For each listener group, we report the per-trial difference,
its 95\% interval and permutation $p$-value, and the per-listener advantage upper bound before and after
block centering.}
\label{tab:measurefull}
\begin{tabular}{@{}llrrlrlrrr@{}}
\toprule
Service & Added & Trials & A's shift, ms & A's $\hat\eps$ & Listeners & Trial diff., ms & $p$ & Raw & Centered \\
\midrule
Anthropic & 131K & 400 & $+2{,}069$ [$+2{,}001$, $+2{,}134$] & 0.963 / 0.987 & own & $-8.4$ [$-44.5$, $+35.3$] & 0.700 & 0.059 & 0.049 \\
 & & & & & other ws & $+12.8$ [$-44.3$, $+42.2$] & 0.666 & 0.088 & 0.071 \\
Anthropic & 98K & 400 & $+1{,}441$ [$+1{,}394$, $+1{,}469$] & 0.983 / 0.996 & other ws & $-7.4$ [$-33.9$, $+22.3$] & 0.713 & 0.062 & 0.054 \\
 & & & & & other org & $+2.9$ [$-16.1$, $+22.9$] & 0.769 & 0.065 & 0.064 \\
OpenAI & 98K & 400 & $+381$ [$+344$, $+425$] & 0.793 / 0.846 & own & $-9.9$ [$-27.7$, $+23.1$] & 0.506 & 0.061 & 0.053 \\
 & & & & & other org & $+10.1$ [$-21.5$, $+37.4$] & 0.554 & 0.074 & 0.055 \\
Gemini & 98K & 384 & $+759$ [$+734$, $+789$] & 0.992 / 0.998 & own & $-1.8$ [$-13.8$, $+10.6$] & 0.739 & 0.069 & 0.060 \\
 & & & & & other proj. & $+10.1$ [$-10.5$, $+22.2$] & 0.258 & 0.065 & 0.050 \\
Anthropic, repeated & 131K & 3{,}000 & $+1{,}405$ [$+1{,}395$, $+1{,}419$] & 0.997 / 0.999 & other ws & $+2.2$ [$-1.9$, $+5.5$] & 0.262 & 0.021 & 0.019 \\
OpenAI, repeated & 16K & 1{,}800 & $+40.3$ [$+27.5$, $+53.0$] & 0.115 / 0.165 & own & $+1.9$ [$-3.3$, $+7.3$] & 0.513 & 0.019 & 0.017 \\
Gemini, repeated & 16K & 1{,}664 & $+95.0$ [$+81.5$, $+109.5$] & 0.425 / 0.462 & other proj. & $+0.6$ [$-8.1$, $+9.1$] & 0.853 & 0.016 & 0.018 \\
Anthropic, prefix & 98K & 195 & $+1{,}654$ [$+1{,}383$, $+1{,}771$] & 0.764 / 0.847 & own & $+11.6$ [$-26.3$, $+45.2$] & 0.451 & 0.088 & 0.078 \\
OpenAI, prefix & 98K & 196 & $+417$ [$+383$, $+484$] & 0.899 / 0.955 & own & $+4.1$ [$-20.9$, $+27.2$] & 0.747 & 0.138 & 0.129 \\
Gemini, prefix & 98K & 196 & $+697$ [$+652$, $+728$] & 0.966 / 0.992 & own & $-7.5$ [$-30.5$, $+15.5$] & 0.352 & 0.070 & 0.045 \\
Anthropic & 10$\times$4K & 400 & --- & --- & other ws & $+5.7$ [$-14.4$, $+21.5$] & 0.573 & 0.036 & 0.038 \\
Anthropic & 16K & 300 & $+338$ [$+268$, $+392$] & 0.486 / 0.613 & other ws & $+11.0$ [$-9.0$, $+27.7$] & 0.262 & 0.046 & 0.047 \\
 & 131K & & $+2{,}543$ [$+2{,}179$, $+2{,}615$] & 0.949 / 0.986 & other ws & $+8.2$ [$-13.3$, $+20.9$] & 0.298 & 0.039 & 0.049 \\
\bottomrule
\end{tabular}
\end{table*}
\noindent\textbf{Design.}
The test sends listeners in three ways, which we call the burst, spaced, and shared-prefix \emph{families},
and we call the computation of A's long prompt the \emph{added work}.
Each burst trial used 21 streamed requests with fresh random text: four listeners preceded requester~A,
and 16 followed within about 12\,ms, or 18\,ms on OpenAI. A sent 64 tokens in the base condition and
about 98{,}304 or 131{,}072 in the added-work condition, and we randomized the two conditions within each
block. In the repeated runs, we sent the earlier listeners about 10 to 2\,ms before A and the later
listeners from about 13\,ms after A, with a median delay of 21 to 32\,ms to the last listener.
The repeated Anthropic run used 131{,}072-token prompts and 1{,}500 trials per condition, whereas the
OpenAI and Gemini runs used 16{,}384-token prompts and 900 and 832 trials per condition, respectively.
OpenAI cached each added prompt, but because every request used distinct content and a distinct cache
key, none read cached tokens. Anthropic refused A's added prompt in three trials, which we excluded from
A's shift and from the per-trial statistics in Table~\ref{tab:lvapi}.

Requests asked for at most 16 output tokens, or one on Anthropic, and outside the shared-prefix trials
none read cached tokens. The per-trial reading uses the median first-token time of the 16 later
listeners, and we compare the two conditions with 10{,}000 permutations and a bootstrap 95\% interval.

The per-listener reading instead uses each later listener as one probe. For it we report cross-validated
single-probe advantage and its bootstrap upper bound, splitting and resampling by trial, before and after
centering each block. The spaced design used 20 listeners from a different Anthropic workspace. They were
spaced 15\,ms apart when A sent ten 4{,}096-token prompts together or sent nothing, and 150\,ms apart when A
sent 64, 16{,}384, or 131{,}072 tokens.

The shared-prefix design tested cache-affinity routing, with A and all 20 listeners sharing a prefix.
Each trial cached 1{,}280 fresh random tokens with an initial request, or 2{,}048 tokens in an explicit
Gemini cache entry, and all burst requests used one key. Listeners extended the prefix by 64 tokens,
whereas A extended it by 64 tokens or to 98{,}304 tokens in total, and usage fields confirmed that every
request in the burst read the prefix. On OpenAI, each such request runs on a machine holding that
prefix~\cite{openai2026prompt}, whereas the other providers do not document whether requests using the
same entry run on the same machine. Each service ran 196 trials in 14 blocks, except that one Anthropic
trial ended in a refusal before the burst, leaving 195. On OpenAI, the raw and centered per-listener
estimates were 0.042 and 0.050, with upper bounds of 0.138 and 0.129, while the per-trial reading showed
no detectable difference. Subtracting the median of each trial's four earlier listeners gave an estimate
of $-0.012$ and an upper bound of 0.084.

These designs cover the burst, spaced, and shared-prefix families of the load-visibility test, and
Supplement~\ref{app:lv} interprets the burst measurements. Every pooled reading in Table~\ref{tab:lvapi}
and every centered pooled reading in the prefix rows fell within its noise bound, the 95th percentile
of the upper bound under the null. For the per-trial statistics, the null distribution of
Section~\ref{sec:evalsetup} gives noise bounds of 0.186 to 0.198 at about 200 trials per condition and
0.074 to 0.093 in the repeated runs. For the per-listener statistics, we instead permuted whole trials
within blocks, which preserves the correlations among listeners in a trial. Their noise bounds were
0.055 to 0.072 for bursts, 0.084 to 0.144 for the shared prefix, and 0.0275 to 0.0281 for repeated bursts.

On Anthropic, most variation in the timing of the 16 later listeners occurred within each burst, with a
robust standard deviation of 100 to 190\,ms. The listeners returned in nearly random order even with
concurrent sending, which cancels network jitter~\cite{vangoethem2020timeless}. Because these providers
control routing, any colocation of a listener with A is incidental. The published burst attack,
by contrast, used listeners on the probe's instance, a quiet queue, and timing that resolved one
step~\cite{prior2026savings}.

\noindent\textbf{Practice.}
We measured only our own requests on our own accounts, and fresh random text kept other users' data and
cache entries out of the experiment; we did not attempt to create or detect cross-account hits.
The runs in Table~\ref{tab:measurefull} cost about \$141 at list prices, mostly on Anthropic, while the
earlier OpenAI runs fell within the free daily allowance.

\section{Throughput Averaged Over Load}\label{app:perf}
\begin{table}[t]
\centering\small
\setlength{\tabcolsep}{3pt}
\caption{Mean throughput across nine load stages on $M$ instances, relative to unprotected sharing. Self and
Cross are the self-reuse and cross-reuse workloads, and Cached is the share of prompt tokens that
unprotected sharing serves from cache; runs with probes omit isolation by account. $^\dagger$The trace
lacks account identifiers, so every hit counts as cross-account and isolation by account coincides with no
reuse.}
\label{tab:perf}
\begin{tabular}{@{}lccc@{\hspace{6pt}}ccc@{}}
\toprule
 & \multicolumn{3}{c}{No probes, two instances} & \multicolumn{3}{c}{Self-reuse, with probes} \\
\cmidrule(lr){2-4}\cmidrule(l){5-7}
 & Self & Cross & Mooncake$^\dagger$ & $M{=}2$ & $M{=}4$ & $M{=}8$ \\
\midrule
Cached      & 0.618 & 0.568 & 0.288 & 0.50--0.51 & 0.420 & 0.366 \\
By account  & 0.945 & 0.802 & 0.720 & ---   & ---   & --- \\
Prediction  & 0.984 & 0.977 & 0.984 & 0.967 & 0.950 & 0.957 \\
Execution   & 0.480 & 0.518 & 0.770 & 0.737 & 0.766 & 0.801 \\
No reuse    & 0.381 & 0.434 & 0.720 & 0.600 & 0.665 & 0.704 \\
\bottomrule
\end{tabular}
\end{table}
Table~\ref{tab:perf} averages throughput across load stages without an SLO. Each cell averages three
runs, except the two-instance runs with probes, which pool four seeds, and the execution mode's cells for
the cross-reuse workload and the Mooncake trace, with one run each, and for four instances, with two.
With probes, the execution mode's throughput is relative to unprotected sharing in the same job. Across
the three load blocks of the self-reuse workload, the prediction mode retained 1.001, 0.984, and 0.966 of unprotected
throughput. On the same workload, the execution mode recovered 0.33 of unprotected sharing's GPU-time
savings over the no-reuse baseline, measured per completed prompt token.
Because the runs on four and eight instances interleave independent copies of the workload
(Supplement~\ref{app:evaldetail}), reuse does not become more available with scale, while more turns run
on a different instance than their predecessors. Unprotected sharing therefore serves fewer tokens from
cache, and the execution mode and the no-reuse baseline gain relative to it.
With probes on 16 and 32 emulated instances, the prediction mode retained 0.956 and 0.962 of
unprotected throughput, while unprotected sharing cached 0.334 and 0.316 of prompt tokens, respectively.
The 32-instance run covered only the middle load block. Figure~\ref{fig:cdf} separates first-token
latency by cache-hit status under unprotected sharing.

\noindent\textbf{Latency and open-loop readings.}
In the closed-loop runs without probes, the prediction mode changes the 90th-percentile time to first
token relative to unprotected sharing by 0.04 to 0.25\,s across stages on the self-reuse workload, 0.05 to
0.21\,s on the cross-reuse workload, and $-$0.05 to 0.10\,s on the Mooncake trace. On the self-reuse
workload, isolation by account adds 0.01 to 0.15\,s, the execution mode 1.2 to 2.8\,s, and the no-reuse
baseline 1.2 to 3.8\,s. At saturation, the prediction mode's throughput exceeds that of isolation by
account by 4.1\% and 21.8\% on the self-reuse and cross-reuse workloads.

In the open-loop runs of Figure~\ref{fig:goodput}, holding at light load raises the 90th-percentile time
to first token by 0.28 to 0.39\,s on the cross-reuse workload and by 0.10 to 0.12\,s on the Mooncake
trace, where unprotected sharing serves 0.22 to 0.30 of prompt tokens from cache. Relative to unprotected
goodput at the same SLO, the prediction mode delivers 0.87 to 0.94 at SLOs of 0.75 to 1.5\,s on the
cross-reuse workload and 0.94 to 0.99 at 0.75 to 2.75\,s on the Mooncake trace. The no-reuse baseline
delivers at most 0.53 on the cross-reuse workload and 0.76 to 0.80 up to 1.75\,s on the Mooncake trace.

\section{Evaluation Details}\label{app:evaldetail}
\noindent\textbf{Implementation.}
The gateway predicts twin steps from its routing records, reports the twin's usage fields, and uses a
dedicated timer to release each first token. It rounds the matched length down to a multiple of 128
tokens, so a request recomputes at most 127 cached tokens.

\noindent\textbf{Workloads.}
The self-reuse workload has 40{,}000 requests with a mean prompt length of 3{,}680 tokens. In it, 10\% of
requests start a conversation, usually with one of the account's
system prompts, and 59\% extend an existing conversation. Another 20\% ask about a document used by
other accounts and 10\% are fresh, and the same generator produces the cross-reuse workload, which shifts
reuse from conversations to such documents. Unprotected sharing serves 0.618 of prompt tokens from cache on
the self-reuse workload and 0.568 on the cross-reuse one, against 0.597 and 0.461 for isolation by account.
For four and eight instances, we interleave independent copies of the self-reuse workload, one pair of
copies per two instances.

\noindent\textbf{Load.}
The routing rule minimizes pending prefill work plus the request's token count minus a weighted cached
match length. The closed-loop client runs nine stages in three groups, which we call \emph{load blocks}, and on two instances its mean
number of requests in flight ranges from 12.3 to 38.6. The open-loop client sends the next turn of a
conversation after the preceding turn's first token arrives. Its nine stages offer 0.40 to 1.04 times
the saturated throughput of unprotected sharing, except that the execution mode and the no-reuse baseline
use their own saturated throughput as the reference. For the Mooncake trace on eight instances, this
reference is not measured: it is four times the two-instance rate scaled by 0.973, the per-instance
ratio between eight and two instances under the no-reuse baseline on the self-reuse workload. The client stops if the queue holds 15\,s of work
or a request has waited 30\,s. After a 40\,s warm-up, stages last 200\,s, except that security runs use
longer stages for the two modes and Mooncake stages last 115\,s. The execution mode's minimum dispatch
delays are 0, 500, and 250\,ms in the three load blocks.

\noindent\textbf{Security runs.}
On two instances, we pooled four seeds, with 8{,}033 to 9{,}017 probes per case for the two modes and
1{,}517 to 1{,}578 for unprotected sharing, which ran in the same jobs as the execution mode. On four and
eight instances, we pooled two and three seeds, with 2{,}075 to 2{,}825 and 6{,}088 to 7{,}927 probes per
case, respectively (Table~\ref{tab:secfull}). The no-reuse baseline,
which has no hits, gave upper bounds of 0.043 and 0.047 with fewer probes. Usage fields followed the
billing rule: no probe reported cached tokens in either mode, whereas under unprotected sharing on two
instances 2{,}703 of the 3{,}107 hit-case probes across both lengths reported them.

The runs with 16{,}384-token hits differ from these in three ways. They cover only the three stages of the
lightest load block, interleave one probe trial per 64 other requests, and raise the engines' maximum
context length to 40{,}960 tokens. Each job ran unprotected sharing, the prediction mode, and the execution
mode in turn. We pooled two seeds on two instances and ran one seed on four and on eight instances, the
latter on a rented machine with eight L40S GPUs, with 114 to 277 probes per case. No probe reported cached
tokens in either mode, whereas under unprotected sharing 394 of the 400 hit-case probes did, and the other
six ran on another instance than their victim. In two of the four unprotected runs, the engines started
with KV pools of 140{,}736 or 146{,}432 tokens. The noise bounds below 250 probes per case come from the
same simulation as the others.

We also examined only the probes whose twin ran on the holder, so that the hit and its twin shared one
instance (Section~\ref{sec:execsec}). On two instances, with 4{,}158 to 4{,}700 such probes per case,
the execution mode's estimates were $-0.004$ and 0.011 for 2{,}048- and 7{,}168-token hits, and their
upper bounds, 0.024 and 0.030, were below the noise bounds of 0.043 and 0.040. On eight instances, with
420 to 456 probes per case, the estimates were 0.014 and 0.087 and the upper bounds 0.100 and 0.171
against a noise bound of 0.138, so the reading for 7{,}168-token hits lay above its bound. With each
probe centered on its 30\,s block, the eight-instance readings became $-0.021$ and $-0.021$, with upper
bounds of 0.085 and 0.066, whereas the two-instance reading for 7{,}168-token hits became 0.024, with an
upper bound of 0.048 that lay above its bound of 0.040. Each of the two readings above its bound thus
returned below it under the other treatment of load.

In the prediction mode, 0.4 to 0.6\% of all hit-case probes used a cross-instance hit and became ready
more than 20\,ms after the target release.

\noindent\textbf{An attacker in the victim's account.}
These runs used the self-reuse workload and the closed-loop stages of the other two-instance runs, and only
the probe's account changed. Each configuration ran once. Isolation by account had 1{,}614 to 1{,}701
probes per case with the probe in the victim's account and 548 to 598 with the probe in another account, where its upper bounds were 0.066 and 0.071 for
2{,}048- and 7{,}168-token hits, within the noise bound of 0.116.
Unprotected sharing, measured alongside, had 1{,}388 to 1{,}458, and the prediction mode had 1{,}408 to
1{,}472.
With the probe in the victim's account, the upper bound of isolation by account for 7{,}168-token hits
was 0.263, beside 0.285 for unprotected sharing. Its advantage was 0.025 below that of unprotected
sharing, and the 90\% interval of their difference, $-0.059$ to 0.016, lies inside the noise bound of
0.082, so within this precision it leaked as much as unprotected sharing. The prediction mode's upper bounds were 0.044 and 0.069
for 2{,}048- and 7{,}168-token hits.
For 2{,}048-token hits, unprotected sharing gave an upper bound of 0.041, and isolation by account gave
0.055 with the probe in the victim's account, so neither detected these short hits. Usage fields confirmed
the setup. With the probe in the victim's account, 1{,}666 of 1{,}701 hit-case probes of 7{,}168 tokens
reported cached tokens, whereas no probe reported them with the probe in another account or in the prediction mode. A second
unprotected run, measured alongside the prediction mode, gave upper bounds of 0.069 and 0.289 for 665 to 718
probes per case.

\noindent\textbf{Simultaneous and repeated probes.}
Even simultaneous probes encounter cache pollution (Section~\ref{sec:probes}), for two reasons. The engine
caches blocks at scheduling time, so only the first probe on each instance computes the prefix. In
addition, our router records the prefix at dispatch, as do SGLang, vLLM's cache-aware router, and llm-d's
approximate index, so later probes favor the first probe's instance even before its blocks enter the cache.

In our light- and full-load trials, eight fresh-account probes for an uncached victim prefix reached
the gateway within a median span of 5 to 14\,ms. Even so, in at least 99.6\% of trials the number of
probes that computed the prefix equaled the number of instances reached, which averaged 1.57 to 1.94 on
two instances and 2.11 to 3.11 on four. Under unprotected sharing, no later probe exceeded the first
probe's advantage by more than 0.01. For 7{,}168-token hits on two instances, the first probe's advantage
was 0.658 and the remaining seven ranged from 0.477 to 0.654. On four instances, the first probe's
advantage was 0.401 and the remaining seven ranged from 0.206 to 0.399.

Because each probe in the security runs has its own victim and prefix, pooling $k$ probes from one case
models repetition without cache pollution. After centering by the median of each load stage, we tested
the mean, median, minimum, and maximum of such sets for $k=1$ to 64. Across both modes, 7 of 363
cells exceeded their noise bound. Among cells with at least 800 sets per case, 2 of 135
exceeded their bounds, compared with 6 of 134 under the no-reuse baseline, where the true effect is zero.
On two instances, both modes' point estimates stayed at or below 0.028 for $k=2$ to 8, and on four and
eight instances the largest upper bound among these cells was 0.080. Under unprotected sharing, by
contrast, the advantage for 7{,}168-token hits rose from 0.41--0.49 with one probe to 0.91--0.97 with 32.

For larger $k$ than the runs resolve, we estimated sensitivity by shifting measured hit-case times and
resampling. A 20\,ms shift required 218 to 1{,}325 probes to reach advantage 0.5, and this count grows as
the square of the noise-to-shift ratio. At each arm's 95\% upper bound on the shift, 10.8 to 82.4\,ms,
the counts ranged from 50 to 865. Under unprotected sharing, the measured gains for 2 to 8 probes never
exceeded the predictions from independently resampling single probes. Other shapes consistent with the
data could, however, require fewer probes: 17 to 112 for a separate hit-case tail, and 6 to 15 for a
monotone likelihood ratio at the upper bound on single-probe distance. All these counts exceed the 1.57
to 3.11 useful probes per burst observed on average before eviction, at most one per instance reached.

\section{Estimating Larger Clusters}\label{app:scale}
The estimates in Section~\ref{sec:evalscale} start with a population of 1.28 million requests drawn as in
the self-reuse workload, of which an $M$-instance cluster serves a random fraction $M/4{,}096$. This coverage
determines the hits available when a request arrives.

An event-level simulator implements the gateway's routing, release, and sharing rules, substitution,
and vLLM scheduling. It runs synthetic requests with these hits on 16 to 256 instances using a
population 16 times smaller. Each run therefore represents the coverage of a cluster 16 times larger,
with full coverage at 256 simulated instances. The simulator matches the two-instance cached shares and the
throughput of the no-reuse baseline within 0.01, and its execution throughput matches the 8- and
32-instance emulated runs within 0.02.

We calibrate models of routing, cross-instance sharing, and substitution on these simulations and
interpolate within the simulated range of $H$. The models overestimate simulated throughput by at most
about 5\%, with the largest differences at the highest $H$. They use the margin required by the
prediction mode's measured cross-instance check at 14.17 requests in flight per instance, whereas the
simulator uses a fixed 100\,ms margin, which is stricter at high reuse. At $H=0.62$, about 2{,}048
instances, the prediction mode reaches 1.38 to 1.40 times the throughput of the no-reuse baseline and the
execution mode 1.23 to 1.25 times. In each range, the lower endpoint comes from the simulator and the
upper endpoint from the models.

\noindent\textbf{Limits.}
We measure security on real clusters of up to eight instances (Section~\ref{sec:evalsec}), and the simulator
estimates throughput only. The estimates have three limits: we validated the simulator up to 32 emulated
instances, we calibrated routing at one load, and the estimates depend on cache capacity. With the
testbed's per-instance KV pool and first-in, first-out eviction, the share available for reuse at
4{,}096 instances falls to 0.55, compared with 0.76 under an unlimited cache. On 32 emulated instances with
probes, the prediction mode's single-threaded gateway uses 0.47 of a core, or 3.8\,ms per request, so a
larger deployment would shard the gateway.

\section{Coverage and Interpretation of the Load-Visibility Test}\label{app:lv}

\noindent\textbf{Which listeners a test must cover.}
For a given route, the second bound of Theorem~\ref{thm:lv} covers the listeners' full observations,
whereas the test measures selected statistics. Each statistic's distinguishing advantage is therefore a
lower bound on load visibility for that route. Using added work as a proxy for a hit or miss also
requires the routing and workload premise in Section~\ref{sec:worlds}.

\emph{Timing on one instance.}
Consider one work-conserving instance that processes prefill requests one at a time in arrival order,
at unit rate, where each request's work is independent of the instance state. We compare two systems
with the same arrivals and work, except that one receives additional work $b>0$ at time~$\tau$.
Let $U_1(t)$ and $U_0(t)$ denote unfinished work with and without this addition, and set $\Delta=U_1-U_0$.

\begin{lemma}\label{lem:drain}
For $t \ge \tau$, $\Delta(t)$ is nonincreasing from $b$, with $\Delta(t) \ge b - (t - \tau)$, and once it
reaches zero the two systems coincide.
A request that arrives at $t > \tau$ starts service, and so returns its first token, exactly $\Delta(t)$
later with the added work.
\end{lemma}

\noindent
After $\tau$, each arrival increases both workloads by the same amount, and between arrivals each workload
falls at unit rate until empty. Thus $\Delta$ starts at $b$ and decreases, at unit rate, only when
$U_0=0<U_1$. Until the two systems coincide, $\Delta$ therefore equals $b$ minus the time the system
without the added work has spent idle since $\tau$. When $\Delta$ first reaches zero, both systems are
empty and receive identical arrivals thereafter. Finally, under first-come-first-served scheduling, an
arriving request waits for exactly the work it finds unfinished.

On one instance, the listener closest behind the added work therefore experiences the largest and most
frequent delay, and if it sees no delay, neither does any later listener. This justifies the burst
family's timing in this model, in which the earlier listeners serve as unaffected references. In a
batched engine, however, an unfinished earlier listener may share a step with the added work, and the
reference measurements check for this effect.

\begin{table*}[t]
\centering\footnotesize
\setlength{\tabcolsep}{4pt}
\caption{Commercial API burst measurements, pooling both listener groups, with about 200 trials per
condition, or 832 to 1{,}500 in repeated runs. For one listener we report the advantage upper bound after
block centering, the concentration function $C$ of first-token time at A's shift, and the estimated
probability $p$ of a shift at least that large. For all 16 later listeners we give advantage estimates
and upper bounds for three trial statistics: Median uses the later listeners alone, while Median diff.\
and Mean diff.\ subtract the median or mean of the four earlier listeners from that of the later
listeners. Mean shift gives the upper bound on mean listener shift relative to A's own shift.}
\label{tab:lvapi}
\begin{tabular}{@{}lrrrrrrrrr@{}}
\toprule
 & & & \multicolumn{3}{c}{One listener} & \multicolumn{3}{c}{All 16 listeners} & \\
\cmidrule(lr){4-6}\cmidrule(lr){7-9}
Service & Added & A's shift & Upper & $C$ & $p$ & Median & Median diff. & Mean diff. & Mean shift \\
\midrule
Anthropic & 131K & 2{,}069\,ms & 0.049 & 0.978 & 0.050 & $-0.028$ / 0.111 & $-0.048$ / 0.121 & $-0.013$ / 0.132 & 2.4\% \\
Anthropic & 98K  & 1{,}440\,ms & 0.039 & 0.992 & 0.039 & $-0.014$ / 0.118 & 0.040 / 0.158 & 0.030 / 0.146 & 3.2\% \\
OpenAI    & 98K  & 381\,ms     & 0.037 & 0.847 & 0.044 & $-0.051$ / 0.120 & 0.026 / 0.168 & $-0.023$ / 0.110 & 9.8\% \\
Gemini    & 98K  & 759\,ms     & 0.035 & 0.995 & 0.035 & $-0.028$ / 0.111 & $-0.036$ / 0.135 & 0.002 / 0.148 & 5.1\% \\
Anthropic, repeated & 131K & 1{,}405\,ms & 0.019 & 0.997 & 0.019 & $-0.011$ / 0.050 & $-0.005$ / 0.047 & 0.008 / 0.051 & 0.6\% \\
OpenAI, repeated    & 16K  & 40\,ms      & 0.017 & 0.166 & 0.100 & $-0.010$ / 0.056 & $-0.014$ / 0.062 & $-0.012$ / 0.058 & 51.5\% \\
Gemini, repeated    & 16K  & 95\,ms      & 0.018 & 0.355 & 0.051 & $-0.012$ / 0.055 & $-0.014$ / 0.057 & $-0.002$ / 0.064 & 6.0\% \\
\bottomrule
\end{tabular}
\end{table*}

\emph{Timing on several instances.}
Now let $M$ identical instances serve requests as above, with the gateway routing each arrival to the
instance with the least unfinished work. Let $\mathbf{U}_1(t)$ and $\mathbf{U}_0(t)$ be the sorted
workload vectors with and without the added work.

\begin{lemma}\label{lem:multi}
For $t \ge \tau$, $\mathbf{U}_1(t) \ge \mathbf{U}_0(t)$ in every component, so every later request routed by
this rule waits at least as long with the added work.
The total excess $\sum (\mathbf{U}_1 - \mathbf{U}_0)$ is nonincreasing from $b$, and once it reaches zero the
sorted vectors, and with them the waits of all later requests routed by this rule, coincide.
\end{lemma}

\noindent
The first claim follows from the monotonicity of the Kiefer--Wolfowitz workload-vector
recursion~\cite{kiefer1955theory}. Adding work to any one instance cannot decrease any sorted component.
Between arrivals, each component decreases at unit rate toward zero, which preserves componentwise order,
and at an arrival, adding the same work to the smallest component and sorting again also preserves it.
For the second claim, arrivals raise the two totals equally, while between arrivals each total falls at
a rate equal to its number of busy instances. Because the system with added work has at least as many
busy instances, the excess cannot grow, and equal totals with ordered components imply equal vectors.

A listener sent $t$ after the added work can be delayed only if total excess is positive at $\tau+t$,
and the probability of that event cannot increase with $t$. The same monotonicity implies that excess is
nondecreasing in the amount of added work, as the premise of Section~\ref{sec:worlds} requires when
routing is the same.

On several instances, however, the closest listener need not experience the largest delay. With two idle
instances, the added work can occupy one while the first listener runs without delay on the other. If a
background request then arrives after that listener finishes but before the added work completes, both
instances are busy and the next listener waits, whereas without the added work one instance would be idle.
The spaced family covers these later arrivals.

\emph{Routing.}
Gateways may route by a cached prefix or a cache key, and the shared-prefix family tests this affinity
with a client-owned prefix. A real probe, by contrast, begins with the guessed prefix, so a listener that
shares it is itself a probe. Unless clients control a separate routing key, the test thus gives listeners
more control over routing than the threat model assumes.

\emph{Limits.}
Both lemmas assume that other arrivals do not respond to the added work. In a closed loop, however, added
work can delay a client's next request, so excess work need not drain and a later listener may even be
faster. The test counts shifts in either direction and does not require either lemma. Its coverage, however,
excludes attackers who time trials to service states more finely than we sampled or who observe more
state than listener timing reveals. In the execution mode, a hit can also change which queued requests
serve as substitutes (Section~\ref{sec:execcluster}), a change in service order that
Theorem~\ref{thm:lv} covers. Because added work that misses draws no substitutes, however, the test's
premise need not hold for that change.

\noindent\textbf{From a reading to the chance of a shift.}
We couple the same trial with and without the added work and let $s_1$ and $s_0$ be the values of a
scalar listener statistic in these trials, such as one listener's first-token time.

\begin{proposition}\label{prop:conv}
Suppose that the added work never lowers the statistic, $s_1 \ge s_0$, and let $E$ be the event that it
raises it by at least $\delta$, $s_1 \ge s_0 + \delta$.
Then the load visibility of the added work is at least $\sup_t \Pr[E,\ t - \delta < s_0 \le t]$.
If moreover $E$ is independent of $s_0$, it is at least $\Pr[E]\,C(\delta)$, where
$C(\delta) = \sup_t \Pr[t - \delta < s_0 \le t]$ is the concentration function of the statistic without
the added work.
\end{proposition}

\noindent
For any $t$, $\Pr[s_0 \le t] - \Pr[s_1 \le t] = \Pr[s_0 \le t < s_1] \ge \Pr[E,\ s_0 \le t < s_0 + \delta]$, and
the left side is at most the total variation distance.

Under these assumptions, the probability $p$ of shifting a listener by at least $\delta$ is at most
its load visibility divided by $C(\delta)$. If the same bound holds for each of $K$ listeners, a union
bound then gives at most $Kp$ for a shift of this size in any listener.

For the commercial APIs, we set $\delta$ to the observed shift in A's first-token time and compute
$C(\delta)$ from trials without added work. Dividing the per-listener reading by $C(\delta)$ then yields
estimates of 0.019 to 0.100 for $p$ (Table~\ref{tab:lvapi}). This interpretation assumes that added work
never makes a listener faster and that the event of a shift of at least $\delta$ is independent of
baseline listener time, and neither assumption is established for these services. Keeping only the first
assumption, Markov's inequality bounds the same probability by the mean listener shift divided by
$\delta$, which gives 0.6\% to 51.5\%. Choosing $\delta$ from A's delay also assumes that A's time grows
with its work. Finally, these estimates concern shifts of at least $\delta$, whereas Theorem~\ref{thm:lv}
counts changes of any size.

  \FloatBarrier
  \ifIncludeMain\else
    \bibliographystyle{plain}
    \bibliography{bib/refs}
  \fi
\fi
\end{document}